\documentclass[11pt]{article}
\pdfoutput=1
\usepackage{graphicx}
\usepackage{hyperref}
\usepackage[inline]{enumitem}
\usepackage{caption} 
\usepackage{lmodern,mathtools}
\usepackage{booktabs}
\usepackage[round,sort]{natbib}
\usepackage{amsmath,amssymb,amsbsy,amstext, amsthm, simplewick}
\usepackage{comment}
\usepackage[margin=1cm,labelfont={sf,bf,scriptsize},textfont={sf,scriptsize}]{caption}
\usepackage{setspace}

\newcommand{\be}{\begin{equation}} \newcommand{\ee}{\end{equation}}
\newcommand{\bea}{\begin{equation} \begin{aligned}} \newcommand{\eea}{\end{aligned} \end{equation}}
\newcommand{\code}[1]{\texttt{#1}}

\newenvironment{alphafootnotes}
{\par\edef\savedfootnotenumber{\number\value{footnote}}
	
	\setcounter{footnote}{0}}
{\par\setcounter{footnote}{\savedfootnotenumber}}

\makeatletter
\newlength{\apb@width}
\newcommand{\autoparbox}[2][c]{\settowidth{\apb@width}{#2}\parbox[#1]{\apb@width}{#2}}

\makeatother

\begin{document}

\begin{titlepage}

\setcounter{page}{1} \baselineskip=15.5pt \thispagestyle{empty}

\vbox{\baselineskip14pt
%\hbox{}
}

%\bigskip\
\vspace{-2.5cm}
\begin{center}
{\fontsize{19}{36}\selectfont  
{
Comprehensive Probabilistic Tsunami Hazard Assessment in the Makran Subduction Zone
}

}
\end{center}

%\vspace{0.6cm}

\begin{center}
\begin{alphafootnotes}
	Parastoo Salah\footnote{\code{parastoo.salah@s.k.u-tokyo.ac.jp}}$^{,1}$,
	Jun Sasaki,$^2$ Mohsen Soltanpour$^3$
\end{alphafootnotes} 
\end{center}

\begin{center}
\vskip 8pt

\textsl{
\emph{$^1$Graduate Program in Sustainability Science-Global Leadership Initiative, Graduate School of Frontier Sciences, The University of Tokyo, Kashiwa, Chiba, 277-8563, Japan,}}
\vskip 7pt
\textsl{\emph{$^2$Department of Socio-Cultural Environmental Studies, Graduate School of Frontier Sciences, The University of Tokyo, Kashiwa, Chiba, 277-8563, Japan}}
\vskip 7pt
\textsl{\emph{$^3$Department of Civil Engineering, K. N. Toosi University of Technology, No. 1346, Vali-Asr St., Tehran, Iran}}

\end{center}

\vspace{0.3cm}
\hrule \vspace{0.1cm}
\vspace{0.2cm}
{ \noindent \textbf{Abstract}
\doublespacing
After the 2004 and 2011 tsunamis came unprecedented to the scientific community the role of probabilistic tsunami hazard assessment (PTHA) in tsunami-prone areas came to the fore.
The Makran subduction zone (MSZ)  is a hazardous tsunami-prone region;
however, due to its low population density, it is not as prominent in  literature.
In this study, we assess the threat of tsunami hazard posed to the coast of Iran and Pakistan by the MSZ and present a comprehensive PTHA for the entire coast regardless of population density.
We accounted for sources of epistemic uncertainties by employing event tree and  ensemble modeling.
Aleatory variability was also considered through probability density function.
Further, we considered the contribution of small to large magnitudes and
used our event trees to create a multitude of scenarios as initial conditions.~\code{Funwave-TVD} was employed to propagate these scenarios.
Our results demonstrate that the spread of hazard curves for different locations on the coast is remarkably large, and the probability that a maximum wave will exceed 3 m somewhere along the coast reaches $\{13.5, 25, 52, 74, 91\}$ for return periods $\{50,100, 250, 500, 1000\}$, respectively. Moreover, we found that the exceedance probability could be higher at the west part of Makran for a long return period, if we consider it as active as the east part of the MSZ. Finally, we demonstrated that the contribution of aleatory variability is significant, and overlooking it leads to a significant hazard underestimation, particularly for a long return period.

\vspace{0.4cm}

 \hrule

\vspace{0.2cm}}
\end{titlepage}

%\tableofcontents

%%%%%%%%%%%%%%%%%%%%%%%%%%%%%%%%%%%%%%%%%%%%%%%
%%%%%%%%%%%%%%%%%%%%%%%%%%%%%%%%%%%%%%%%%%%%%%%
%%%%%%%%%%%%%%%%%%%%%%%%%%%%%%%%%%%%%%%%%%%%%%%
\section{Introduction}
\label{sec:Intro}
Tsunami events are infrequent in several water bodies around the world, yet their danger cannot be ignored due to the high levels of destruction that follow, including major losses of life and property damage. In particular, the importance of a comprehensive tsunami hazard assessment (THA) is highlighted when a disastrous tsunami occurs. Recent devastating tsunamis such as the Sumatra tsunami of 2004, with more than 200,000 fatalities \citep{cultureTsunami}, and the 2011 Tohoku tsunami in Japan, which caused more than 15,000 fatalities and was responsible for the Fukushima Nuclear Power Plant accident~\citep{2011tsunami}, are representative examples. Following these disasters, there has been a remarkable development in tsunami risk management to reduce the effect of future tsunamis. For recent reviews of these developments, including full lists of references, see \citep{tsunamireview2,riskreview}.

Tsunami hazard assessment includes sensitivity analyses (see e.g. \citep{slip3,sensitivity4}) as well as deterministic (see e.g. \citep{deterministic2,deterministicmakran,salah}) and probabilistic approaches. The latter approach-- called the probabilistic tsunami hazard assessment (PTHA) --has received substantially increased attention after the 2004 and 2011 tsunamis \citep{surprise1,mmaxjapan,surprise3,deterministic}.
Unlike deterministic approaches that consider specific scenarios (commonly including the worst case scenario) to calculate tsunami hazard metrics (such as run up height and arrival time), PTHA calculates the likelihood of tsunami impact employing multiple possible scenarios consisting of the contributions from small to large events along with all quantifiable uncertainties~\citep{definition}. Hence, PTHA can overcome the limitation of incomplete or insufficient historical records, and extend the return periods from hundreds to thousands of years. Furthermore, this approach considers the uncertainties stemming from the lack of researcher knowledge and the random nature of hazards. The former is represented by epistemic uncertainty in literature while the latter is known as aleatory variability. These concepts are explained in detail in section~\ref{sec:uncertainty}.

PTHA was developed by adopting the probabilistic seismic hazard analysis (PSHA)~\citep{psha,ptha,ptha2}, and much progress has been built upon it see~\citep{review} and the references therein. Notwithstanding that PTHA is a relatively new method, it has been widely used in tsunami-prone areas owing to its diverse range of applications (e.g.,~\citep{EUSA,Eaustralia,Ejapan}), each of them covers different uncertainties, methods, and level of accuracy.
In this study, we assessed the tsunami hazard using the probabilistic approach in  the Makran subduction zone (MSZ).

The MSZ is a tsunami risk zones as attested by compiled tsunami catalogues and recent paleotsunami studies~\citep{makranunesco} that exhibits risks for the neighboring countries of Iran, Oman, and Pakistan. This region is not as prominent in scientific literature as other tsunami-prone subduction zones owing to its low population density, and it remains as one of the least studied regions. The authors of~\citep{heidarzadeh2011probabilistic} performed the first generation of PTHA in the MSZ.
Their results are not reliable for return period far from the typical recurrence time of magnitude $M_\text{w}=8.1$ because only three earthquakes were considered in their study. Furthermore, the rough discretization of sources used may have affected the final results.
~\citep{makranprobabilistic} conducted a PTHA along the MSZ based on a synthetic earthquake catalogue. In their study, a simple geometry model along with a uniform (cf. heterogeneous) slip distribution were used because their primary focus was identifying the consequences of maximum magnitude assumptions. Finally,~\citep{makran2probabilistic} performed a logic tree approach for assessing the hazard only for Oman coasts.
Of particular importance is the absence of aleatory variability in the aforementioned studies. For any tsunami probability study, it is critical to understand how uncertainty affects probability estimation. Thus, we aim to fill the gaps of previous PTHA studies in the MSZ via developing a methodology that incorporates both aleatory and epistemic uncertainties. Our work overcomes the limitation in the integration of uncertainties, namely, tidal level, heterogeneity in slip distribution and rupture size, numerical and geometry models, earthquake recurrence rate, and maximum magnitude.

First, we quantified the epistemic uncertainties of fault source for the assessment of mean annual rates of earthquakes at different magnitude levels. Despite the more classical approaches commonly used in literature, we employed event tree and  ensemble modeling, which is based on a method initially introduced  for PSHA studies~\citep{ensemble}. To develop our event tree, we utilized available seismic, geodetic, and historical catalogue data to better understand the potential seismogenic zone, maximum magnitude, and recurrence model for the MSZ. Next, rupture complexity, namely, dimensions, slip distribution, and possible earthquake locations, were considered to develop scenarios. Then, a high-resolution tsunami numerical model was used to propagate tsunami waves resulting from these scenarios. Finally, we consider the aleatory variability associated with  tidal variations, tsunami numerical and bathymetric models, and scaling relations through statistical methods. We followed these specific intermediate steps to derive the probability of tsunami height occurrence and exceedance for a given exposure time along the Iran and Pakistan coasts. We also compared our results obtained in the presence and absence of the aleatory variability. Our findings provide information for various stakeholders to underpin tsunami risk  activities, such as insurance activity, land use and city planning, critical facility design, and mitigation measure design and implementation.

\section{Methodology and Dataset}\label{sec:Method}
\begin{figure}[h]
	\begin{center}
		\includegraphics[width=\textwidth]{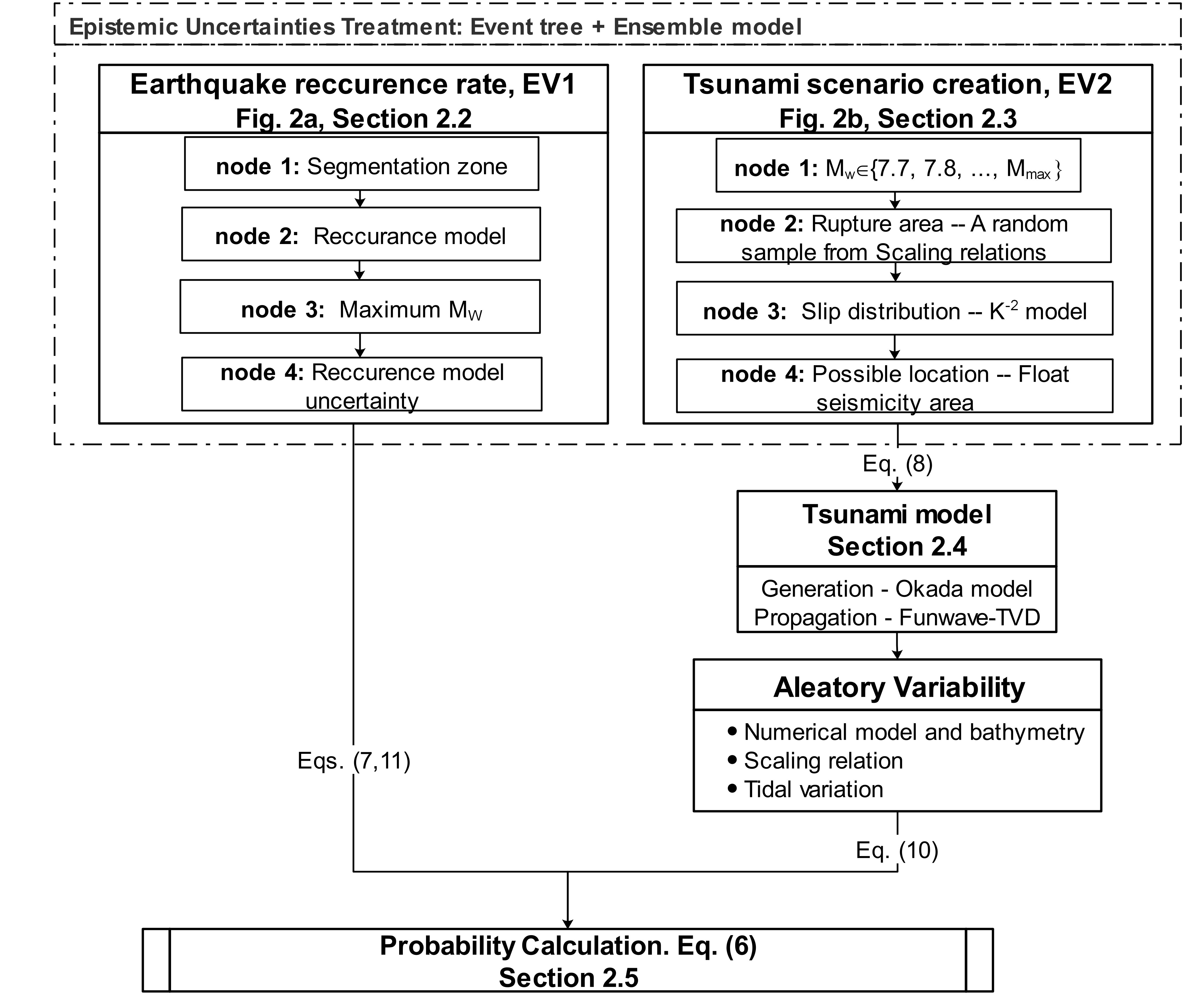}
		\caption{Methodology framework. First, the fault geometry was defined using SLAB 2.0, and the source was discretized into smaller segments. Next, two event trees were developed to define the earthquake recurrence rate and create tsunami scenarios; then, the Okada model and \code{Funwave-TVD} were used to calculate tsunami heights for our scenarios. Finally, considering the aleatory variability, we derived the probability of exceedance.}
		\label{Fig:flowchart}
	\end{center}
\end{figure}

Our methodology aims to calculate the probability of exceeding a set of tsunami heights at the Makran coast, considering both epistemic and aleatory uncertainties. In this work, we only focused on tsunamis induced by earthquakes; landslide-induced tsunamis were beyond the scope of our research and should be addressed in future work. Fig. \ref{Fig:flowchart} demonstrates a summary of our framework. First, we determined the seismicity area and generated synthetic scenarios similar to that described by \citep{Eaustralia}. Then, for each scenario, we ran a fully nonlinear tsunami model \code{Funwave-TVD}~\citep{funwave1,funwave2} to obtain the maximum wave heights along the coastline. Additionally, we incorporated the epistemic uncertainties by developing two event trees and ensemble modeling. Finally, we calculated the tsunami height exceedance rate considering the aleatory variability.

\subsection{Treatment of uncertainties}\label{sec:uncertainty}
A reliable PTHA must consider the epistemic uncertainty and aleatory variability simultaneously. The latter expresses the innate variability of the physical process, while the former is related to the lack of understanding and limited knowledge of the process.\footnote{Many authors believe that no theoretical significance exists for this separation because, as long as our knowledge increases, all uncertainties become epistemic~\citep{ensemble}.} As shown in Fig. \ref{Fig:flowchart}, each factor was considered as described in detail below.
\begin{figure}[h]
	\centering
	\includegraphics[scale=1]{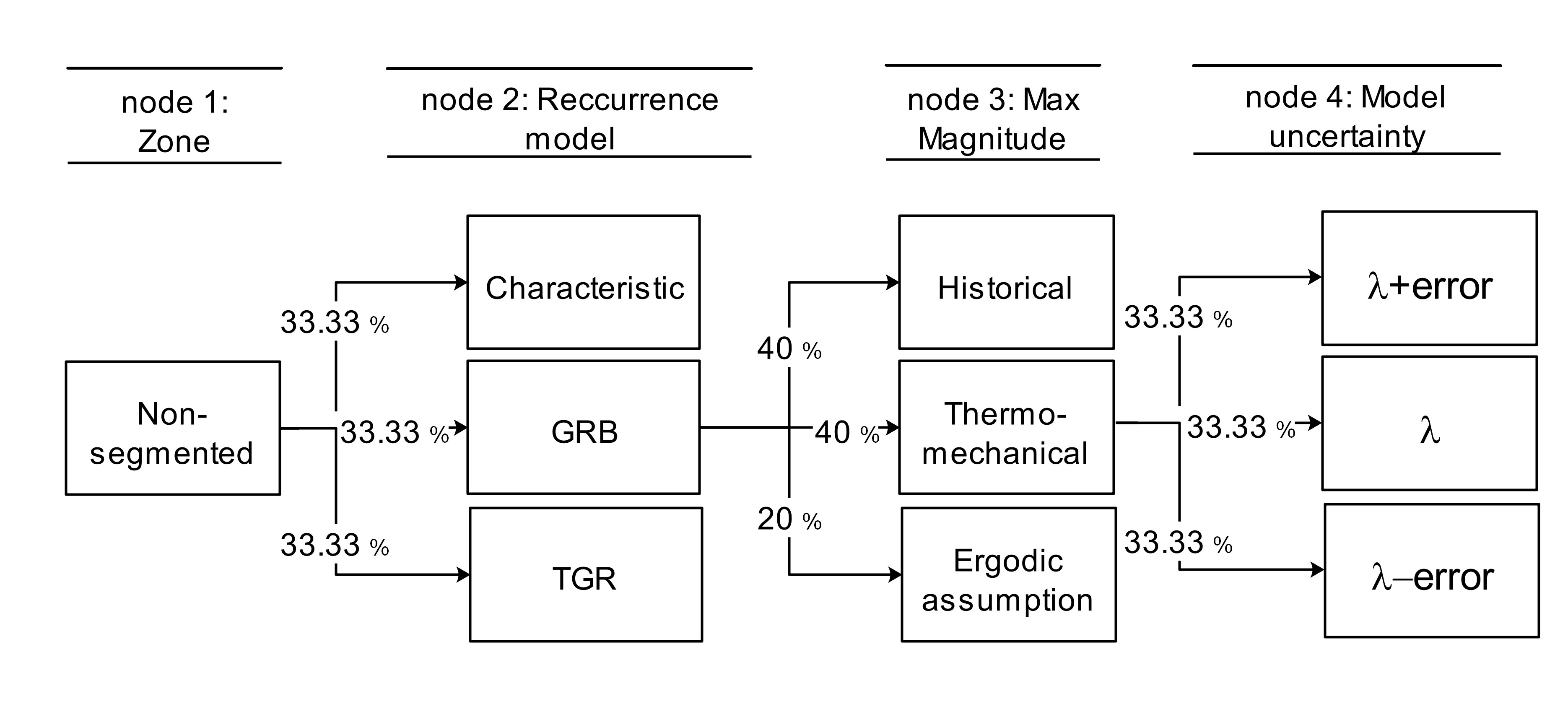}
	\caption*{(a)}
	~\\
	\includegraphics[width=\textwidth]{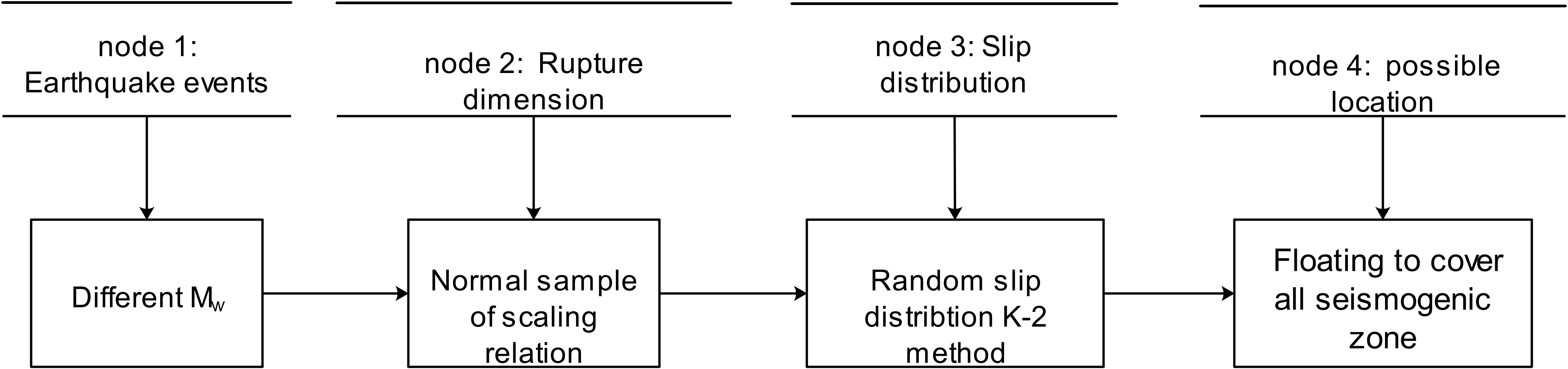}
	\caption*{(b)}
	\caption{Developed event trees for (a) source recurrence model; (b) rupture complexity and tsunami scenario creation.\label{Fig:ev}}
\end{figure}
\subsubsection{Epistemic}\label{Sec:epistemic}
Epistemic uncertainties can be incorporated by developing event trees~\citep{logoctreeAnnaka}. We developed two event trees:
\begin{enumerate}[wide=0pt,widest=\textwidth,label=(\roman*)]
	\item  Focusing on the fault source recurrence model for the assessment of mean annual rates of earthquakes at different magnitude levels with 27 branches. It consists of three approaches for the seismicity model [Gutenberg--Richter--Bayes (GRB) \citep{GRBKijko}, truncated Gutenberg--Richter, and characteristic \citep{characteristicKagan}]; three maximum magnitudes ($M_\text{max}$) [based on the Kijko--Sellevoll--Bayes method~\citep{MmaxKijko}, thermomechanical modeling~\citep{smith2013thermal} and ergodic assumption~\citep{ErgodicAssumption}]; and three for incorporating the uncertainty of the earthquake occurrence model. See Fig. \ref{Fig:ev}
	\item   Focusing on the bulk rupture parameters and rupture complexity. It consists of rupture length and width, earthquake source location within the fault, and slip distribution. See Fig. \ref{Fig:ev}
\end{enumerate}
\subsubsection{Aleatory}\label{Sec:aleatory}
The proper treatment of aleatory variability in  tsunami wave heights is a prominent subject, and ignoring this typically leads to significant hazard underestimation~\citep{Aleatorybommer2006modern}. In our analysis, we have identified three main contributions, i.e., $\{\sigma_{\text{m}}, \sigma_{\text{s}}, \sigma_{\text{t}}\}$, to the aleatory variability as below.

\paragraph*{Numerical model and bathymetry ($\sigma_{\text{m}}$) --}

\begin{figure}[h]
	\centering
	\includegraphics[scale=.4]{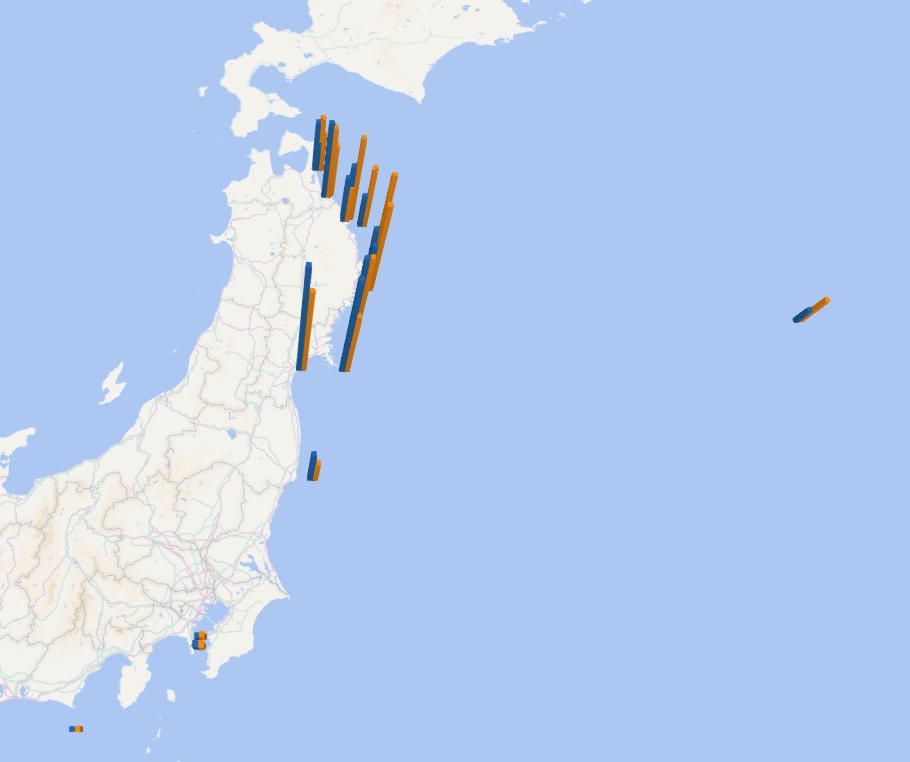}
	\includegraphics[width=8cm]{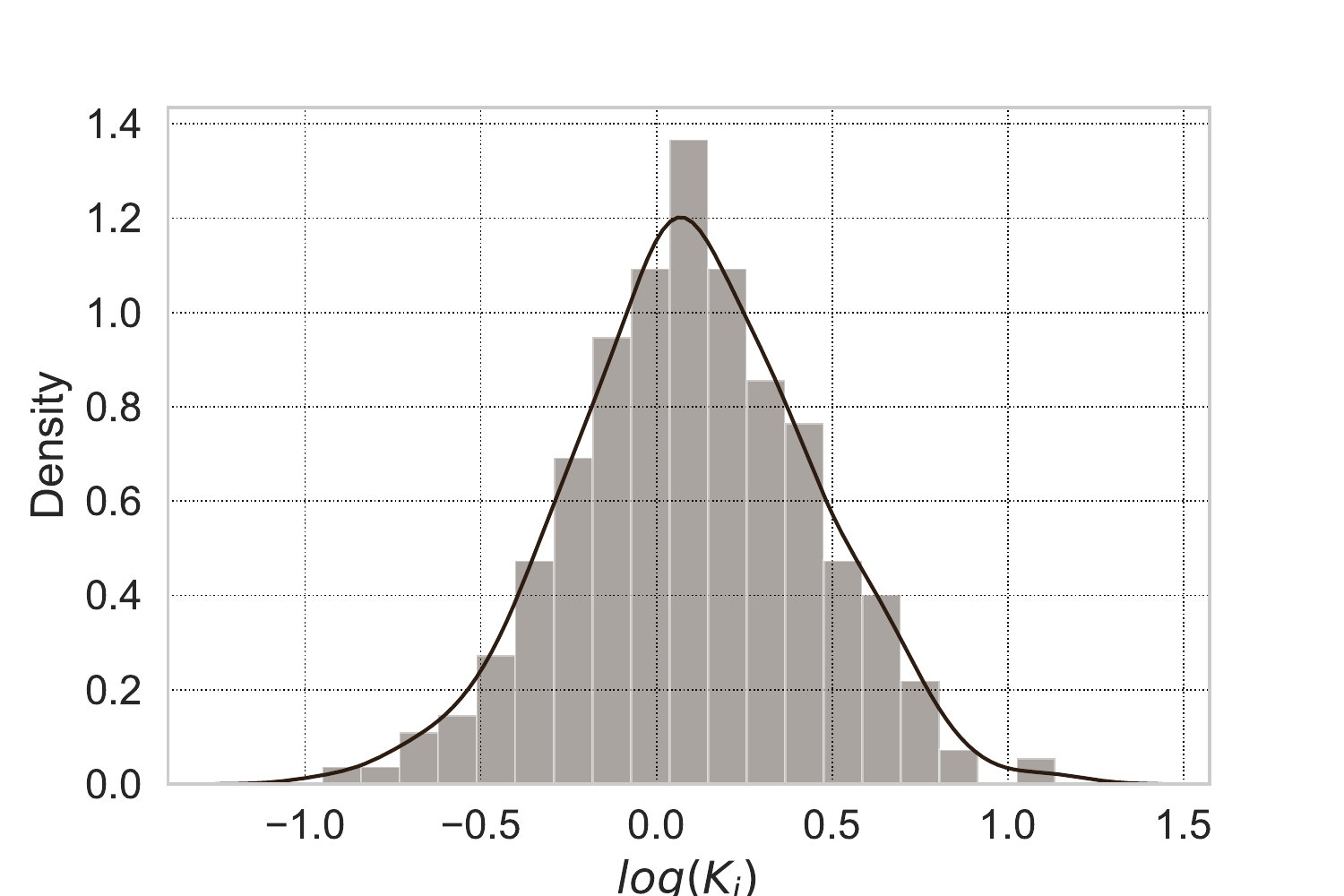}
	\includegraphics[width=6.5cm]{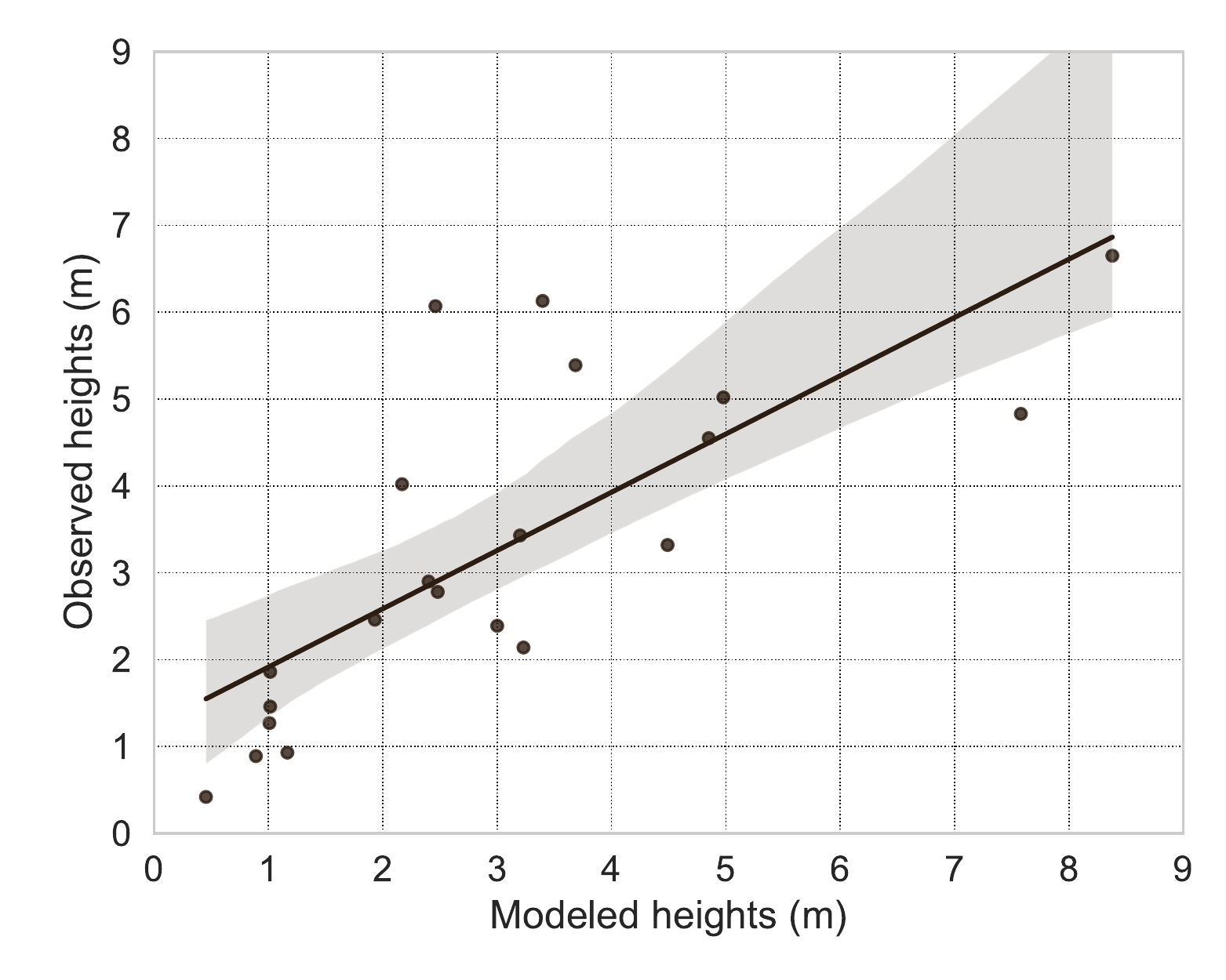}
	\caption{Comparison between modeled and measured tsunami height for the 2011 Japan tsunami at 15 stations recorded by GPS, DART buoys, tide and wave gauges; regression line for modeled versus measured height (bottom right); histogram of errors in log tsunami height and corresponding normal distribution (bottom left). 
		\label{Fig:kappa}}
\end{figure}

~Due to the lack of field data and background information on the MSZ, the 2011 Tohoku earthquake of Japan was modeled, and the results were compared with the available measured data to quantify the mismatch between the observed and computed tsunami heights. This uncertainty is described as the standard deviation of a log-normal distribution with a zero mean~\citep{aida,bathy2}:
\bea
\sigma_{\text{m}} & = \log \kappa = \sqrt{\frac{1}{n} \sum_{i=1}^{n} (\log K_i)^2 - (\log K)^2} \, , \\
\log K & = \frac{1}{n} \sum_{i = 1}^{n} \log\left( \frac{H_{\text{obs, }\! i}}{H_{\text{model, }\! i}} \right) \, .
\eea
Here, $K_i= H_{\text{obs, }\! i}/{H_{\text{model, }\! i}}$ with $H_{\text{obs, }\! i}$ and $H_{\text{model, }\! i}$ being the measured and simulated tsunami heights at the $i^{\text{th}}$ station, respectively.
For $H_{\text{obs}}$, the measured tsunami height at GPS, DART buoys, and tide and wave gauges\footnote{The data from these source were used to avoid uncertainties when using survey measuring methods.} were used. Moreover, we simulated the 2011 Tohoku tsunami using the same bathymetry and numerical model as the ones we used for the MSZ to obtain  $H_{\text{model}}$ (see section~\ref{Sec:tsunami}). Fig. \ref{Fig:kappa} shows the comparison between the modeled and measured tsunami heights with $\sigma_{\text{m}}=0.376$.
\paragraph*{Scaling relations ($\sigma_{\text{s}}$) --}
~Given the earthquake magnitude, rupture length and width were derived by evaluating the scaling relations. To do so, we used Strasser relations \citep{scalingstresser} as explained in section~\ref{Sec:scaling}. To account for stochasticity in the earthquake dimensions imposed by the scaling relations, we used the standard deviations associated with the equations, which were $\sigma_{\text{s}_1}=0.173$ and $\sigma_{\text{s}_2}=0.180$ for length and width, respectively.
The variability in scaling relations was derived from a regression analysis of the relations~\citep[Table 1]{scalingstresser}. $\sigma_{\text{s}} = \sqrt{\sigma_{\text{s}_1}^2+\sigma_{\text{s}_2}^2}$= 0.249 was used in combination with the other sources of aleatory variability (see section \ref{Sec:calculation}).
\paragraph*{Tide ($\sigma_{\text{t}}$) --}
~Because the tide level at tsunami arrival time is unknown, tidal variation variability must be included in the PTHA. In the Makran region, the tidal variation is notable, and the peak-to-peak tidal amplitude is as high as $2$-$3$ m. For this task, we calculated the probability of exceedance of mean sea level (MSL) from the tidal record at each point of interest (PoI).

\begin{figure}
	\includegraphics[width=\textwidth]{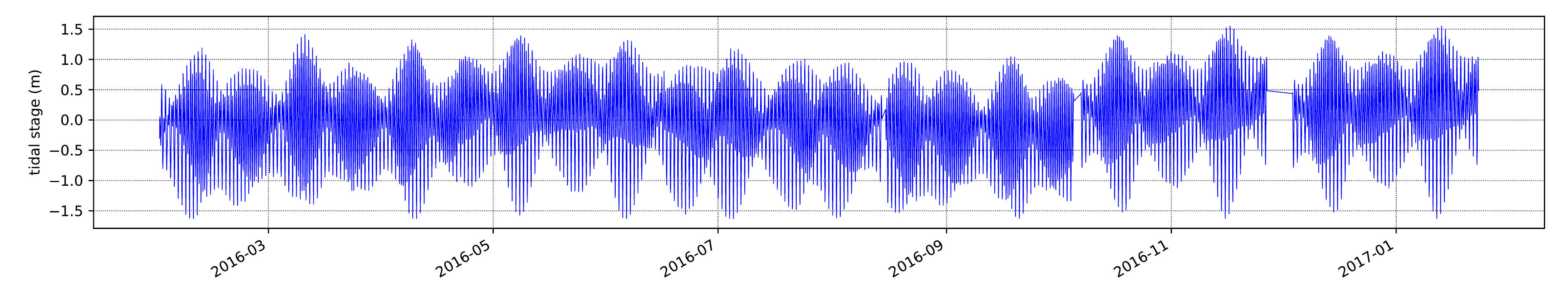}
	\begin{center}
		\includegraphics[scale=.5]{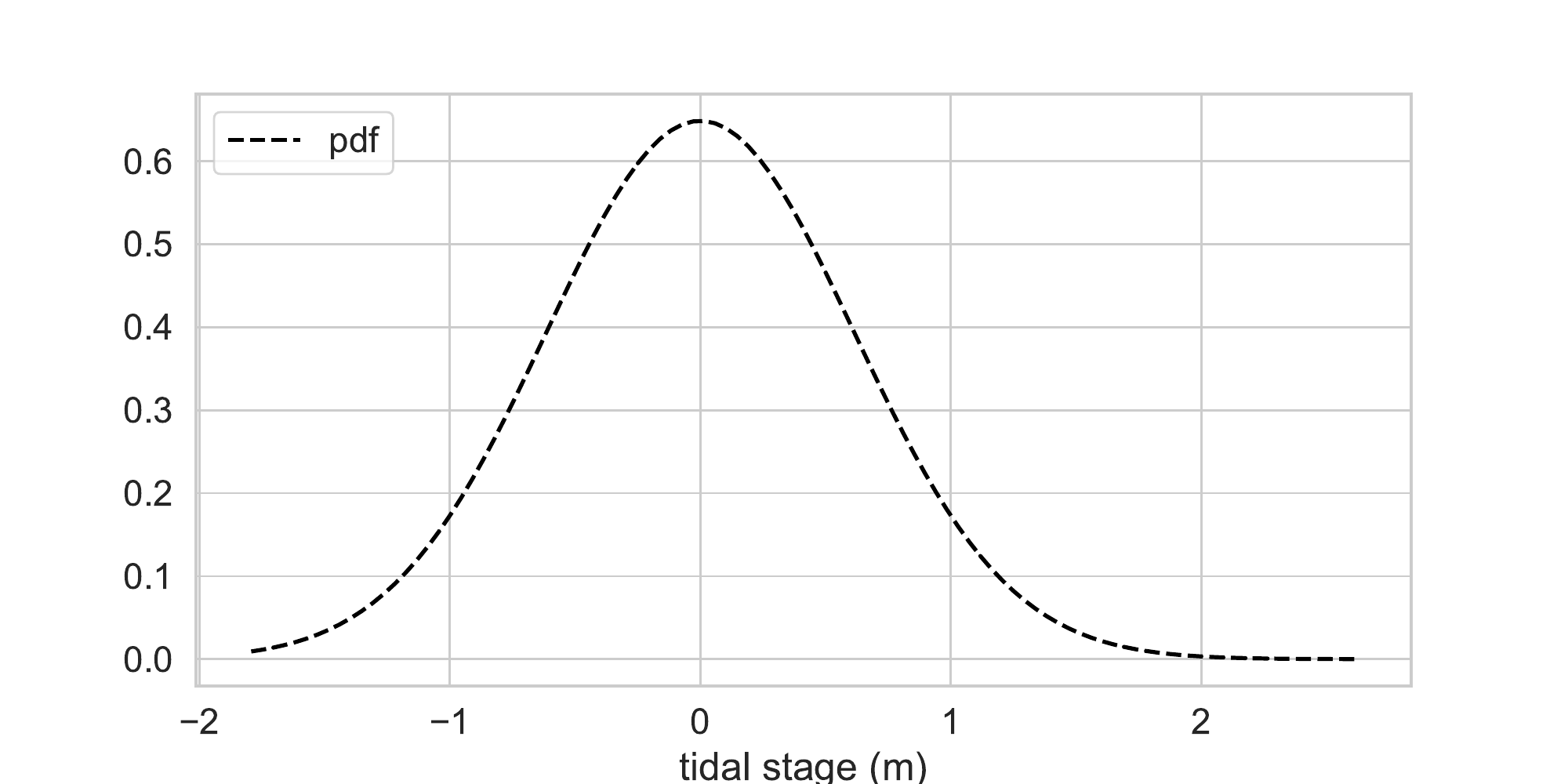}
	\end{center}
	\caption{Tidal time series record of one year starting from 2016 for Beris (top); corresponding normal distribution (bottom). For this PoI, $\sigma_{\text{t}}=0.612$.
		\label{Fig:tide}}
\end{figure}

To calculate tidal record probability, we used a relatively long  time-series of record measured by tidal gauges for each PoI. For PoIs in which a tidal record is not available, we used a linear interpolation of the closest tidal gauges. This choice seems reasonable  because the differences in tidal levels along the Makran coast are not significant~\citep{tidevariation}. Fig. \ref{Fig:tide} illustrates an example of our methodology for one PoI, Beris.
$\sigma_{\text{t}}$ for the remaining  PoIs are presented in Online Resource 1.

\subsection{Source}\label{Sec:EV1}
The MSZ is located on the southeastern coasts of Iran and southern coasts of Pakistan. This zone extends east from the Strait of Hormoz to the Ornch--Nal Fault in Pakistan. It experienced the deadliest tsunami that has occurred in the Indian Ocean prior to 2004, and recent smaller earthquakes suggest seismicity on the megathrust. However, poor historical records have led to significant uncertainty and complicated hazard potential estimation. Therefore, as mentioned  in section~\ref{Sec:epistemic}, to incorporate the uncertainties associated with the fault source, we developed an event tree (EV1) to assess the mean annual rates ($\nu_j$) of  earthquakes at different magnitude levels, as described below.	

\subsubsection{Zone: node 1 in EV1}
The eastern and western parts of the MSZ exhibit extremely different seismicity patterns~\citep{makran}. This, along with its unrecognized bathymetric trench, makes the MSZ a unique subject of analysis.
~\citep{makranseg} argued that the eastern MSZ is underlain by an oceanic lithosphere, while the western part is possibly underlain by a continental or very low velocity oceanic lithosphere. This, along with the more historical seismicity activity at the eastern part, form the hypothesis of east-west segmentation of the MSZ. However, it remains a controversial issue whether the MSZ should be considered segmented in hazard studies
because the existence of late Holocene marine terraces along the eastern and western halves suggests that both can generate megathrust earthquakes~\citep{makranholo}.

 Owing to the above mentioned related controversy, we neglected the hypothesis of the segmented MSZ as it leads to a strong hazard underestimation.~\footnote{Note that treating Tohoku as a segmented zone led to strong underestimation of the devastating 2011 tsunami  ~\citep{mmaxjapan}.} Accordingly, this study considers only the non-segmented rupture.

\subsubsection{Recurrence rate model: node 2 in EV1}
The severity of a large earthquake is determined by the tail of a frequency distribution. Thus, earthquake catalogues are limited at	large magnitudes for a particular fault zone. This makes the accurate estimation of the probabilistic tsunami hazard through the application of the recurrence interval of seismic history impossible. In particular, for MSZ with poor and incomplete catalogues, a simple linear regression of the historical cumulative  distribution is known to be biased~\citep{Newzland}. Accordingly, several models exist that can be used to define the distribution of earthquake magnitudes for incomplete catalogues. In this study, we used three seismicity models:\begin{enumerate}[label=(\roman*)]
	\item Gutenberg–Richter–Bayes(GRB)~\citep{GRBKijko}. Seismicity was determined using the \code{HA3} application built in \code{MATLAB}. The applied procedure of the seismic hazard considers the incompleteness of the seismic catalogues, uncertainty in magnitude estimation, and variation in seismicity. The code accepted mixed data catalogues, namely, paleo, historical, and instrumental  with different completeness magnitudes, time periods, and magnitude uncertainties. This method employed a mixed (Bayesian) Poisson-gamma distribution as a model of earthquake occurrence over time.
	\item Characteristic~\citep{characteristicKagan}. The characteristic distribution has the cumulative complementary function ($\Phi_M$) truncated on both ends and is characterized by the following equation
	\be
	\Phi_M = \Bigg\{
	\begin{array}{lr}
		e^{- \beta ( M - M_\text{min} )} \hspace{0.5cm} \text{for } & M_\text{min}\leq M\leq M_\text{max}\\
		~ & ~ \\
		0    \hspace{2.3cm}\text{for } &  M > M_\text{max}
	\end{array} \, .
	\ee
	
	\item Truncated Gutenberg-Richter (TGR). The cumulative complementary function ($\Phi_M$), which is truncated at both ends, is expressed as
	\be
	\Phi_M = \Bigg\{
	\begin{array}{lr}
		\frac{e^{- \beta ( M - M_\text{min} )}-e^{- \beta ( M_\text{max} - M_\text{min} )}}{1-e^{- \beta ( M_\text{max} - M_\text{min} )}} \hspace{0.5cm} \text{for } & M_\text{min}\leq M\leq M_\text{max}\\
		~ & ~ \\
		0    \hspace{4.54cm}\text{for } &  M > M_\text{max}
	\end{array} \, ,
	\ee
	where $M_\text{min}$ is the level of magnitude completeness, $M_\text{max}$ is the maximum possible earthquake magnitude and $\beta=b\log10$, and $b$ is the parameter of the Gutenberg-Richter relation.
\end{enumerate}
\subsubsection{$M_\text{max}$: node 3 in EV1}
PTHAs are more sensitive to $M_\text{max}$ than PSHAs because tsunami heights do not saturate with increasing magnitude as seismic ground motions do~\citep{bathy2}. $M_\text{max}$ based on instrumental catalogues may underestimate the maximum magnitude event due to their short records. Here, to include this uncertainty, we used three methods for maximum magnitude ($M_\text{max}$) assessment:	 
\begin{enumerate}[label=(\roman*)]
	\item  Kijko-Sellevoll-Bayes method~\citep{MmaxKijko}: using the \code{HA3} application, we found $M_\text{max}=8.2$.
	\item Thermomechanical model: we observed a potential of  $M_\text{max}=9.22$ for the full length of subduction zone in~\citep{smith2013thermal}.
	\item Ergodic assumption: \citep{ErgodicAssumption} suggested $M_\text{max}=9.58$ for subductions based on their  statistical analysis for a number of faults worldwide.
	 However, this is an implausible event in comparison to the aforementioned maximum magnitudes \citep{frohling2016gps}. Thus, a smaller weight (20 $\%$), was assigned to this branch of our event tree.
\end{enumerate}

\subsubsection{Earthquake catalogues}
Earthquake data employed in this study were derived from various sources:
\begin{enumerate*}[label=(\roman*)]
	\item International Seismological Centre (ISC)
	\item Incorporated Research Institutions for Seismology (IRIS)
	\item The United States Geological Survey Online bulletin (USGS), which includes information from the National Ocean and Atmospheric Administration (NOAA) and Preliminary Determination of Epicentres (PDE) provided by the National Earthquake Information Center (NEIC)
	\item Global Historical Earthquake Archive (GEM)
	\item Iranian Seismological Center (IRSC)
\end{enumerate*}.	Extra effort has been made to extract additional data from literature regarding earthquakes with magnitudes beyond 6.5. This includes information from the Pakistan Meteorological Department (PMD)~\citep{PMD} and \citep{persianearthquake}.

We compiled the catalogues for a region that lies in the plate interface, excluding nonsubduction seismicity (see Fig. \ref{Fig:bathy}). The catalogues cover the period from 825 BCE to mid-2020 CE. These catalogues are different in terms of magnitude scale. When available, the moment magnitude, $M_w$,  was used; otherwise, the published magnitudes (e.g., teleseismic magnitudes and modified Mercalli intensity) were converted to $M_w$ using the  empirical $laws$ proposed by~\citep{mmi,msmw}.

We  used the \code{ZMAP7} analysis tool~\citep{zmap} to prepare the catalogues for our recurrence models. First, following the assumption that seismicity obeys a Poisson process,
it is necessary to decluster the catalogues by removing all dependent events, namely, precursors and aftershocks.
Hence, we employed the cluster approach proposed by Reasenberg~\citep{reasenberg} to eliminate dependent shocks. Then, duplicate events from different catalogues were removed. Sunsequently, the plot of the  cumulative number of events allowed us to split the working catalogues into prehistorical, historical, and three sub-instrumental categories. Each has a different magnitude of completeness ($M_c$) and magnitude uncertainty. Moreover, we obtained a \code{prior} value for $b$ in each catalogue to use in our recurrence models (see Table~\ref{table:zmap}).
\begin{table}
\caption{Extracted values for magnitude of completeness ($M_c$), error, and $b$-value from \code{zmap} for different working catalogues.}
	\centering
	\begin{tabular}{ | c || c | c | c | c | c |}
		\hline
		& prehistorical & historical & complete 1 & complete 2 & complete 3 \\ \hline
		period & $326 \text{ BC } - 1020$ AD & $1480-1899$ & $1900-1963$ & $1964-1989$ & $1990-2020$ \\ \hline
		$M_c$ & --- & $5.5$ & $5.7$ & $4.8$ & $4.8$  \\ \hline
		error value & $0.6$ & $0.5$ & $0.45$ & $0.35$ & $0.25$ \\ \cline{1-6}
		prior $b$-value & \multicolumn{5}{ c| }{$0.91 \pm 0.04$} \\ \hline
	\end{tabular} \label{table:zmap}
\end{table}

\subsection{Tsunami scenarios
}\label{sec:scenario}
To create possible tsunami scenarios and incorporate rupture and location uncertainties, event tree 2 (EV2) was developed (see Fig. \ref{Fig:ev}). The branches of EV2 are introduced in section~\ref{Sec:epistemic}; here, we describe them in detail.
\subsubsection{Source discretization} Similar to~\citep{Eaustralia}, fault geometry was defined using a three-dimensional source zone fault-plane, SLAB 2.0 -- a comprehensive subduction zone geometry model~\citep{slab2}. The MSZ has an extremely shallow subduction angle (dip) and thick sediment pile ($\approx 7 $ km) that leads to a wide potential seismogenic zone~\citep{smith2013thermal}. Following the suggestion of \citep{slab} and \citep{makrandepth}, we constrained the seismogenic zone from 0 km (i.e., trench) to $38$ km depth as a preferred down-dip limit. This assumption leads us to define a seismogenic zone for the MSZ as shown in Fig. \ref{Fig:bathy}.

Then, to obtain a better representative of the MSZ fault geometry, we discretized our seismogenic zone into $50\times 50$ km$^2$ segments. Finally, dip, rake, strike, and depth for each segment were identified for use in the Okada model \citep{okada} to generate the initial tsunami conditions. The fault parameters for all the segments are presented in  Online Resource 2.
\subsubsection{Rupture area: node 1 and 2 in EV2}\label{Sec:scaling} For each magnitude ranging from $M_w=7.7$ \footnote{$M_w=7.7$ is the minimum magnitude capable of causing a noticeable tsunami.}  to $M_w=9.5$ with a regular magnitude interval of 0.1, i.e., $M_w \in \{M_{w,\text{min}}, M_{w,\text{min}}+0.1,\ldots,M_{w,\text{max}}-0.1, M_{w,\text{max}}\} $, we calculated the rupture length and width using the scaling relation of Strasser derived from the regression analysis of historical subduction events \citep[Table 1]{scalingstresser}.
For $M_w \leq 8.7$, we included uncertainties associated with the use of the scaling relation for earthquake dimensions as described in section \ref{sec:uncertainty}. However, we observed that the variability enlarges with growing magnitude. Hence, for $M_w >8.7$, rather than using only one value for rupture length and width, a
random sample was selected from a log-normal distribution.  Our initial intention was to consider the dependence of the variance of rupture length and width, and sample from a two-dimensional multivariate normal distribution~\citep{scalingmulti}. However, we noticed that the range of variation was quite small. Considering our segmentation size (i.e., $50 \times 50$ km$^2$), the independent random selection of  length ($L$) and width ($W$) were  generated form normal distributions according to
\be
 \begin{aligned}
&\log_{10} L \sim \mathcal{N}(- 2.477 + 0.585 M_w,0.18) \, , \\ &\log_{10} W \sim \mathcal{N}(- 0.882 + 0.351 M_w,0.173)  \, .
 \end{aligned}
\ee
Here, $\mathcal{N}(\mu,\sigma)$ is a normally distributed random variable with mean $\mu$ and standard deviation $\sigma$; notation $\sim$ denotes the equivalence of distributions.

Then, for each length and width we calculated the number of segments downdip ($ n_s$) and along-strike ($n_l$) using the method described in~\citep[Eq.~(4)]{global}.

    \subsubsection{Slip distribution: node 3 in EV2} Slip distribution significantly affects tsunami heights nearshore. Recently, different studies have shown that maximum nearshore wave height varies by a factor of 2 or more due to heterogeneity in earthquake slip~\citep{slip1,slip2,slip3,slip4,slip5}. However, owing to its convoluted nature and  computation complexity, tsunami hazard assessments are usually based on idealized uniform slip earthquakes. In this work, we used a uniform slip for $M_w \leq 8.7$, where the effect of spatial slip distribution is not significant, and heterogeneous slip distribution for  $M_w \geq 8.8$, where we found that the heterogeneity of slip notably varies tsunami heights at our PoIs. This trade-off was specified to account for the effect on tsunami heights and optimize the number of scenarios through our sensitivity analysis. The number of scenarios and differences among modeled tsunami heights at PoIs were compared for a fixed scenario, but with varying  $M_w$, starting from $M_w=7.7$.

This observation is similar to~\citep{slip4} in which, for the Japan PTHA, earthquakes with $M_w>8.9$ were considered large, and the authors included three levels of spatial slip in their model.

Average slip was computed for each scenario with magnitude $M_w$ employing the scaling relation as follows:
\be \label{Eq:slip}
M_w=\frac{\log M_o -9.1}{1.5} \, ,\qquad \qquad S=\frac{M_o}{\mu \times A} \, ,
\ee
where $M_o$ is the seismic moment, $\mu$ the shear modulus, and $A$ the area of each scenario. We set $\mu=3\times10^{11}$ dyn cm$^{-2}$ as it is appropriate for crustal rocks and shallow depth faults~\citep{omanrock}. We then used the evaluated $S$ from Eq.~\eqref{Eq:slip} as a uniform slip for $M_w \leq 8.7$; whereas for $M_w >8.7$, the slip for each sampled ($L$,$W$)-scenario was created randomly using the \code{PTHA18} code built in \code{R}. The \code{PTHA18} code uses the SNCF model of~\citep{slipmodel} for generating random slip distribution for a given segment dimension and number. This model is a variant on
the widely used $K^{2}$ model. Further implementation details can be found in~\citep{scalingaustralia} and~\citep{slipmodel}.
\begin{figure}
	\hspace{-2.5cm}
	\includegraphics[width=1.22\textwidth]{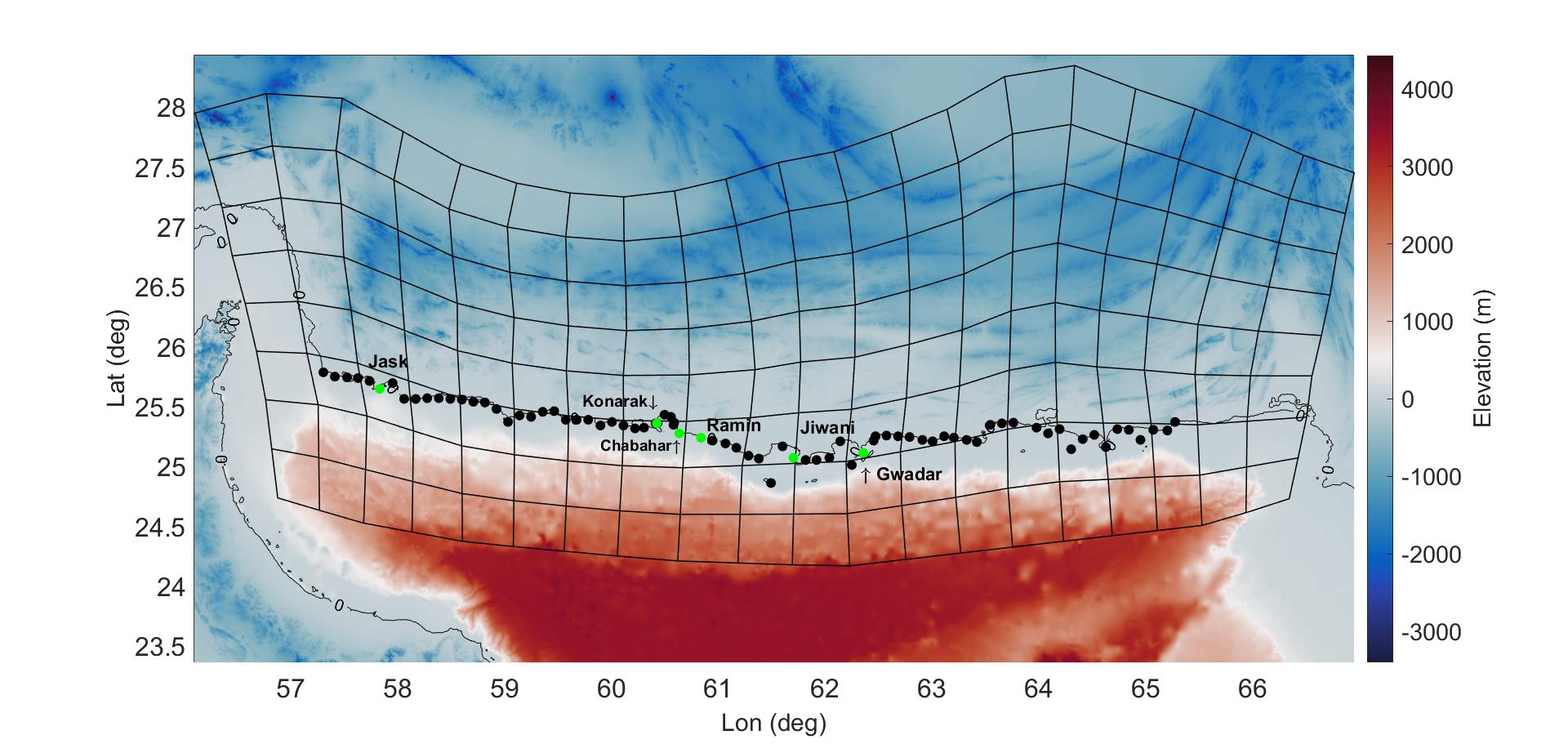}
	\caption{Elevation (m) and computational domain; dots represent PoIs located at 0 m isobath along the coast; black mesh indicates the seismogenic zone and source discretization into $50 \times 50$ km$^2$.}
	\label{Fig:bathy}
\end{figure} 
 
\subsubsection{Possible location: node 4 in EV2}
To cover all the seismogenic zone for each magnitude and sampled length and width, we floated the calculated $ n_s \times n_l$ through all possible locations of the Makran seismicity area, shown in the blue mesh in Fig.~\ref{Fig:bathy}. We assumed that the occurrence of a specific magnitude was equally probable in all possible locations; therefore, an equal weight was assigned to the branches of EV2.
\subsubsection{Analysis of the number of scenarios}
\begin{figure}[!ht]
	\centering
	\includegraphics[scale=.85]{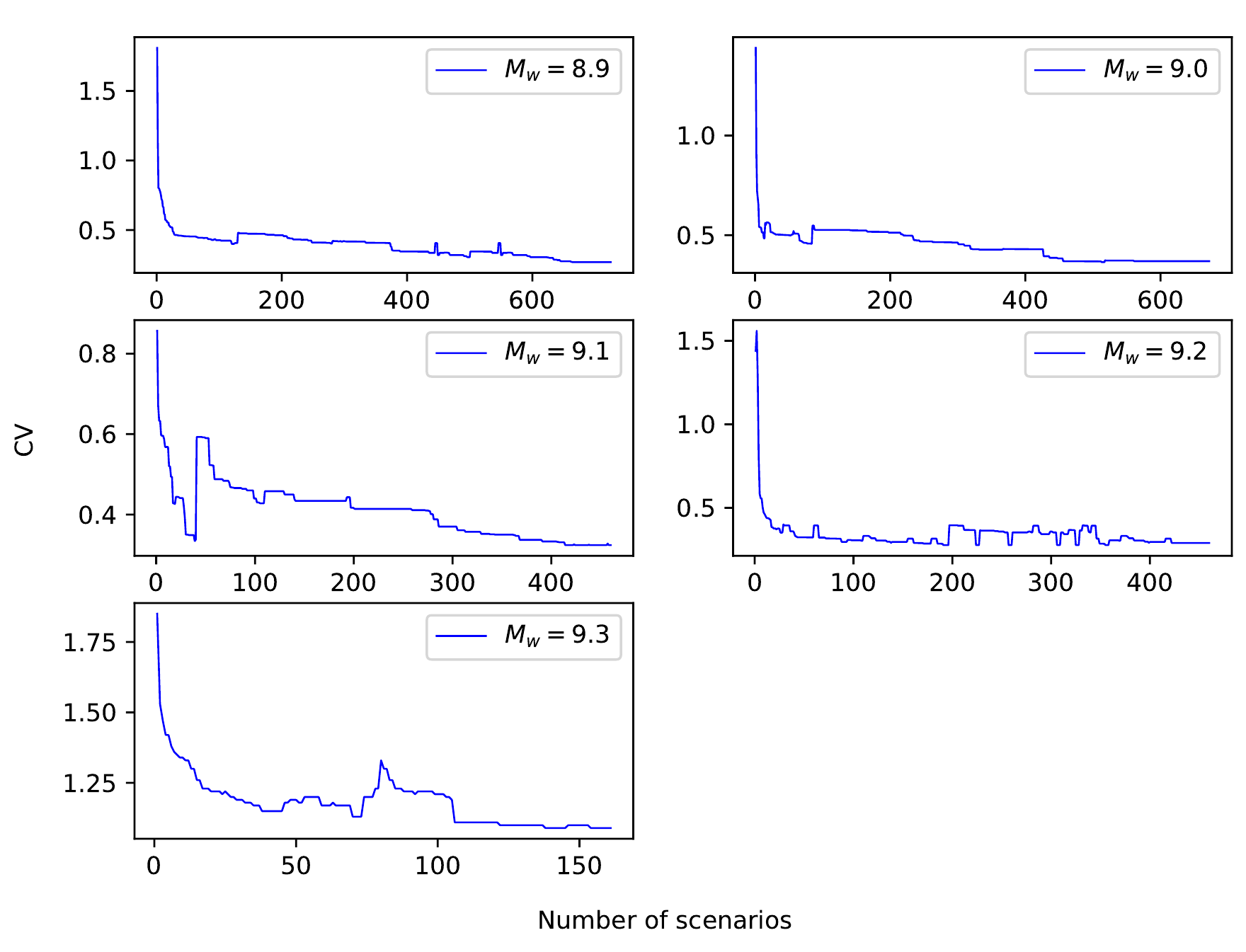}
	\caption{Coefficient of variation of maximum coastal tsunami heights at different level of magnitudes against the number of scenarios.}
	\label{Fig:CV}
\end{figure}

In this section, we describe the number of scenarios based on the method explained in the previous subsections. Moreover, a justification to confirm the sufficiency of the number of scenarios is provided.\footnote{We acknowledge the referee for bringing this point to our attention.}
A uniform slip distribution was used when the difference in the mean value of all the PoIs along the coast was $\leq 0.5$ m. The number of scenarios increases in lower magnitudes (due to the rupture size, thereby floating it to cover all the seismicity areas); however, the tsunami heights at PoIs were not significant, employing heterogeneous distribution. Hence, for  $M_w \leq 8.7$ (which was the threshold to indicate the difference in tsunami heights due to the heterogeneity in slip distribution) we used a uniform slip distribution. 
For $M_w  \geq 8.7$,  a specific number of heterogeneous slip distributions (from 10 to 18) was used. We used the technique introduced in \citep{mulia2020regional} to demonstrate that the number of considered slip distributions for each magnitude is sufficient. Hence, we calculated the coefficient of variation $CV=\sigma_\text{h}/\mu_\text{h}$, where $\sigma_\text{h}$ and $\mu_\text{h}$ are the standard deviation and mean of maximum tsunami heights at all the PoIs, respectively. The results are illustrated in Fig. \ref{Fig:CV}, where the variation of CV approximately converges to zero for the given number of scenarios for each magnitude level. Table \ref{Table:scenarios} shows the number of scenarios for different magnitude levels based on the method described above. \footnote{For $M_w>$ 9.3 the selected value for probability of exceedance (for the longest return period) is zero (see section \ref{epoe})}
\begin{table}[h!]
\caption{Number of scenarios for each magnitude considering the heterogeneity of the slip.}
\label{Table:scenarios}
	\centering
	\begin{tabular} {| p{0.0256\textwidth} | p{0.0236\textwidth}|p{0.0236\textwidth}|p{0.156\textwidth}|p{0.156\textwidth}|p{0.156\textwidth}|p{0.176\textwidth}|}
		\hline
	$M_w$ & $n_s$ & $n_l$ & Possible scenarios along-strike & Possible scenarios down-dip & Total Possible scenarios & Total Possible Scenarios \\ \hline 
		7.7 & 2 & 1 & 19 & 7 & 133&133 \\ \hline
		7.8  & 2 & 2& 19 & 7 & 133&133  \\ \hline
		7.9  & 3 & 2& 18 & 7 & 126&126  \\ \hline
		8.0 & 3 & 2 & 18 & 7 & 126&126 \\ \hline
		8.1  & 4 & 2& 17 & 7 & 119&119  \\ \hline
		8.2  & 4 & 2 & 17 & 7 & 119&119 \\ \hline
		8.3 & 4 & 2& 17 & 7 & 119&119  \\ \hline
		8.4 & 6 & 2 & 15 & 7 & 105&105 \\ \hline
		8.5  & 6 & 3& 15 & 6 & 96&96  \\ \hline
		8.6 & 7 & 3 & 14 & 6 & 84&84  \\ \hline
		8.7 & 8 & 3 & 13 & 6 & 78 &78 \\ \hline
		8.8 & 9 & 3 & 12 & 6 & 72 &72*10 \\ \hline
		8.9 & 11 & 3 & 10 & 6 & 60 &60*12 \\ \hline
		9.0 & 13 & 4 & 8 & 5 & 40 &40*13  \\ \hline
		9.1 & 15 & 4 & 6 & 5 & 30&30*15 \\ \hline
		9.2 & 16 & 4 & 5 & 5 & 25&25*18 \\ \hline
		9.3 & 19 & 5 & 2 & 4 & 8&8*18 \\
		\cline{1-7}
	\multicolumn{5}{| c| }{Sum} & 1428 & 4227 \\ \hline
	\end{tabular}
\end{table}

Alternatively, to justify the number of scenarios, the results of the samples from similar earthquake locations and magnitudes of past earthquake tsunami events can be compared. 
However, notably, there is a significant disagreement between the results of numerical models and existing measurements, evidence, and witnesses for the 1945 tsunami,  which is the only significant tsunami in the MSZ \citep{okal2015field, rastgoftar2016study,heidarzadeh2017combined}. 
This connotes the existence of another mechanism involved in the tsunami generation, such as landslides. Hence, comparing our simulation results with the available data will be neither precise nor beneficial, as we have only considered earthquake-generated tsunamis.
\subsection{Tsunami model}\label{Sec:tsunami}
In total, 4220 scenarios were created using the approach discussed in the previous section considering the branches of EV2 for different magnitudes, which were randomly sampled from the rupture area, slip distribution, and all possible locations. For each scenario, numerical simulations of tsunami generation and propagation were performed. First, we calculated vertical co-seismic dislocation via a homogeneous elastic half-space model~\citep{okada}. Then, the Kajiura filter~\citep{kajiura} was used
for the ocean surface deformation of the dislocation to calculate the initial conditions. Regarding the simulation of tsunami propagation
a fully nonlinear and dispersive Boussinesq long wave model, \code{FUNWAVE-TVD}~\citep{funwave1,funwave2}, was employed. It features accurate dissipation by considering the breaking wave and bottom friction processes, and has been systematically validated against experimental studies and benchmarks~\citep{tehranirad2011tsunami}. The
code was parallelized using the message passing interface (\code{MPI}). This scalable algorithm (using more than $90\%$ of the number of cores in a computer cluster~\citep{funwave2}) has been paved our way for modeling multitude scenarios.
The dispersive effects may not always be significant in long tsunami wave trains and considering them is computationally demanding (cf. nonlinear shallow water models) \citep{tajalibakhsh}. However, owing to the \code{MPI} protocol, using \code{FUNWAVE-TVD} did not cost us notable additional time, given our computational tools.\footnote{The computation was carried out using the computer resource offered under the category of General Projects by Research Institute for Information Technology, Kyushu University, with 36 number of processors.}
Instead,
adopting \code{FUNWAVE-TVD} in this study would be useful for future development of multi-scale PTHA from ocean to local inundation covering most of the important processes with higher accuracy, including landslide induced and far-field tsunamis, where wave dispersivity becomes important.

Here, \code{FUNWAVE-TVD} was used in its Cartesian implementation. To prevent non-physical reflection
from the boundaries, sponge layers were specified with $10$ km thickness within our computational domain. A $600$ m resolution was used for the computational domain, which is a trade-off between precision and practical feasibility (see Online Resource 3).
Moreover, the finer grid resolution would be beneficial only if a high-quality bathymetry is available. However, the high-quality bathymetry data does not exist in almost all the selected PoIs.

Here, GEBCO-2020 (global bathymetric model based on ship-track data) model with 15 arc second resolution, combined with limited measured data in shallow water depths by the Ports and Maritime Organization of Iran (PMO), has been used for our tsunami simulations.
Each scenario has been simulated for $8$ h, and for each computational time step, a time series of tsunami wave has been recorded at $84$ hazard points. These PoIs are located at  $5$ to $0$ m isobath at approximately $10-12$ km intervals along the Iran and Pakistan coastline. The PoIs are shown in Fig. \ref{Fig:bathy}.

\subsection{Deriving the probability of exceedance
}\label{Sec:calculation}
For a given exposure time ($\Delta T$), PTHA was performed by deriving the  exceedance of maximum tsunami height ($\psi$) at each PoI from a threshold value ($\psi_t$). Considering a total of $J$ possible magnitudes, we defined the total probability of exceedance
\be
P^{\text{tot}} ( \psi > \psi_t , \Delta T, \text{PoI} ) = 1 - \prod_{j = 1}^{J} \left( 1 - \mathcal{P} ( E_j , \Delta T ) P ( \psi > \psi_t  | E_j ) \right) \, ,
\ee
where $\mathcal{P} ( E_j , \Delta T )$ is the probability that at least one event ($E_j$) occurs in the return period $\Delta T$. Assuming that the occurrence of earthquakes conforms to a stationary Poisson process with the annual recurrence rate $\nu_j$, it can be assessed as
\be\label{Eq:earthnu}
\mathcal{P} ( E_j , \Delta T ) = 1 - \exp \left( - \nu_j \times \Delta T \right) \, .
\ee
Considering the uncertainties on rupture dimensions, locations, and slip distribution in EV2, each $E_j$ can cause different scenarios ($\mathcal{S}^{(j)}_A$). The probability that tsunami height ($\psi$) exceeds a threshold ($\psi_t$) when the event $E_j$ occurs is then given by
\be
P ( \psi > \psi_t  | E_j ) = \sum_{A = 1}^{A_j} P \big( \mathcal{S}^{(j)}_A | E_j \big) P \big( \psi > \psi_t | \mathcal{S}^{(j)}_A  \big) \, .
\ee
Here, $P \big( \mathcal{S}^{(j)}_A | E_j \big)$ is the probability of occurrence of the scenario $\mathcal{S}^{(j)}_A$, and in the absence of aleatory variability,
\be
P \Big( \psi > \psi_t \Big| \mathcal{S}^{(j)}_A \Big) = \bigg\{ \begin{array}{lr}
	0 , & \psi < \psi_t \\
	1 , & \psi \geq \psi_t
\end{array} \, ,
\ee
While in the presence of the aleatory variability that was discussed in section~\ref{Sec:aleatory}~\citep{thio2010},
\be \label{Eq:aleatory}
P \Big( \psi > \psi_t \Big| \mathcal{S}^{(j)}_A \Big) = 1 - \Phi \left( \log ( \psi_t ) \big| \left[ \log ( \psi_{\mathcal{S}^{(j)}_A})\right] , \sigma \right) \, .
\ee
$\Phi$ is the cumulative distribution function for a log-normal distribution with the mean equal to the modeled tsunami height at each PoI and standard deviation $\sigma$, given value of a $\log(\psi_{t})$. From~\citep{thio2010}, $\sigma $ can be computed by combining our aleatory variability terms $\sigma = \sqrt{\sigma_{\text{m}}^2+\sigma_{\text{t}}^2+\sigma_{\text{s}}^2}$.

\subsubsection{Ensemble model}\label{sec:ensemble}
In this section, we explain how to incorporate the uncertainties from EV1 and obtain $\mathcal{P} ( E_j , \Delta T )$ using the ensemble model~\citep{ensemble}. To calculate the probability that at least one earthquake $E_j$ occurs for the selected $\Delta T$, an event tree was developed as described in section~\ref{Sec:EV1} and Fig. \ref{Fig:ev}. The branches of  EV1 were treated in the framework of ensemble modeling, as introduced in~\citep{ensemble}. Ensemble modeling presumes that epistemic uncertainty is greater than that evaluated by an event tree, and treats the branches of the event tree as an unbiased sample from a parent distribution. This distribution, $f(\theta)$, describes the variable $\theta$ simultaneously considering the aleatory variability and epistemic uncertainty.

In our case, branches of EV1 are a small sample size, and their few probability outcomes can be replaced by a parametric distribution. A natural choice is the beta distribution that is commonly used in hazard literature.
In this case, we set variable $\theta^{(E_j)} =  \mathcal{P} ( E_j , \Delta T )$ so that  the variable will be the hazard curve. Different $\theta^{(E_j)}$ are the branches of EV1 that are now a sample of a Beta $(\alpha,\beta)$ distribution. Parameters $\alpha$ and $\beta$ are related to the average and variance of   $\theta^{(E_j)}$ as
\be \label{eq:beta}
\text{E} [\theta^{(E_j)}] = \frac{\alpha}{\alpha + \beta} \, , \qquad
\text{Var}[\theta^{ (E_j)} ] = \frac{\alpha \beta}{(\alpha + \beta)^2 (\alpha + \beta +1)} \, .
\ee
In our context, $\text{E} [\theta^{(E_j)}]$  and $\text{Var}[\theta^{ (E_j)} ]$ denote, respectively, the weighted average and variance of the exceedance probabilities of the $j$th magnitude for the selected $\Delta T$. Inverting equations \eqref{eq:beta}, we found the parameters of the Beta distribution for each magnitude $j$. Finally, calculating the Beta parameters of the exceedance probability for a set of magnitudes, we plotted the full hazard curve.

\section{Results and Discussion}
In this section, we present the results obtained from the analyses and modeling presented in previous sections for the coastal area of the MSZ. Our main results are presented by earthquake and tsunami probability exceedance curves and tsunami probability maps for the selected return time periods. In this study, we set $\Delta T=\{50,100,250,500,1000\}$ years; each choice interests different stakeholders and provides information on a specific aspect of the tsunami hazard in the MSZ.  We also compared the results obtained in the presence and absence of the aleatory variability.
\subsection{Earthquake probability exceedance curves} \label{epoe}
\begin{figure}[!ht]
	\centering
		\includegraphics[width=7.8cm]{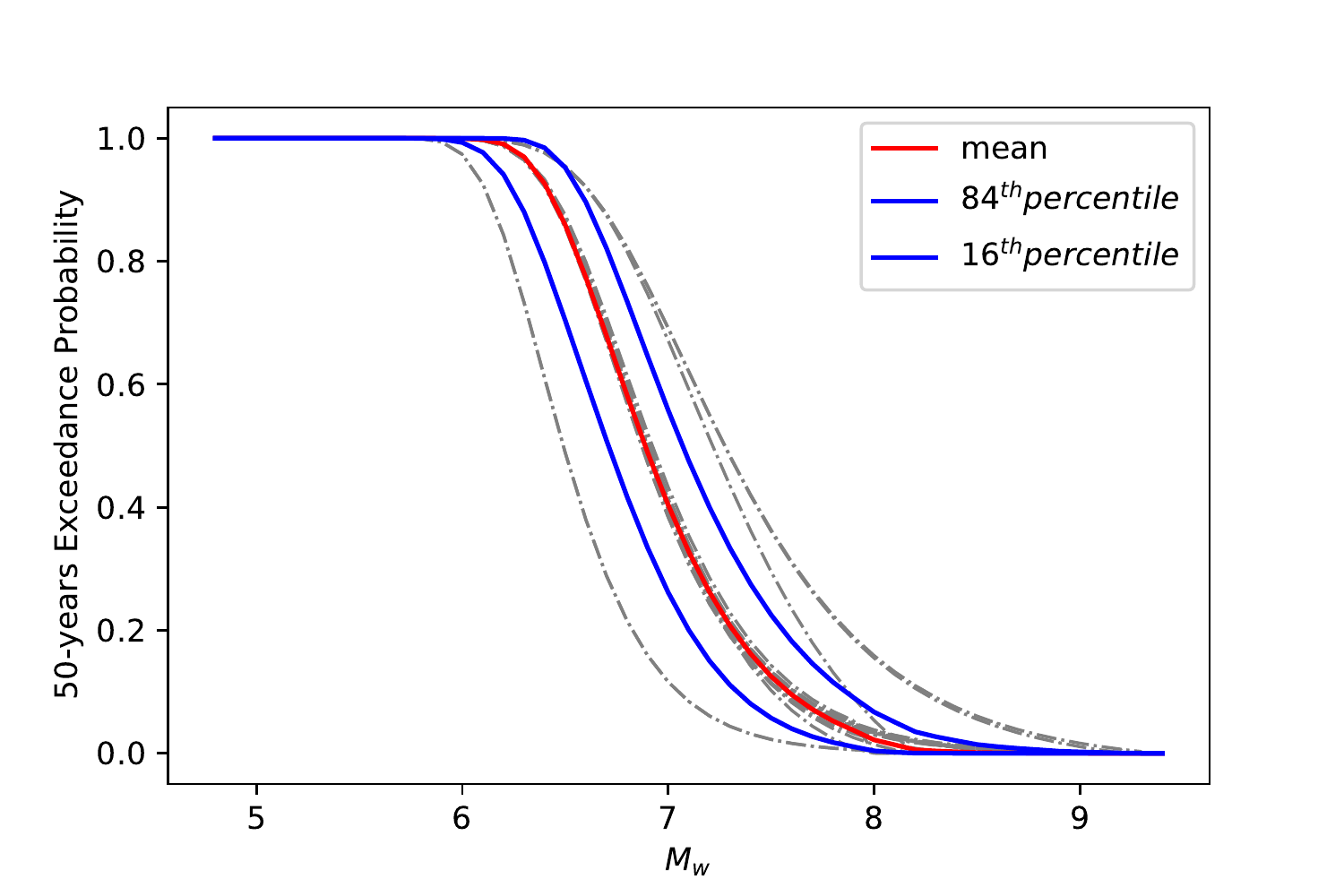}
	\includegraphics[width=7.8cm]{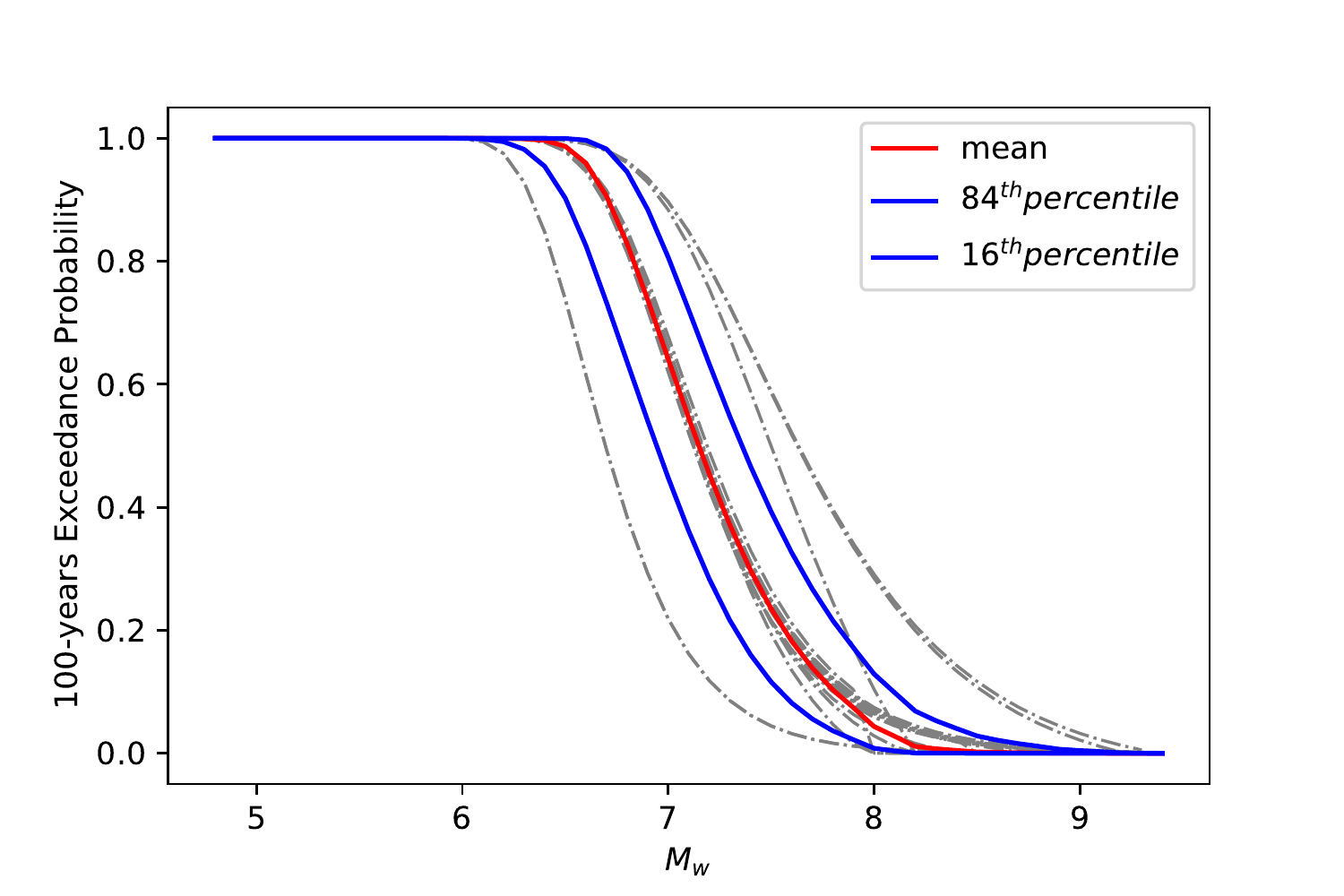}
	\includegraphics[width=7.8cm]{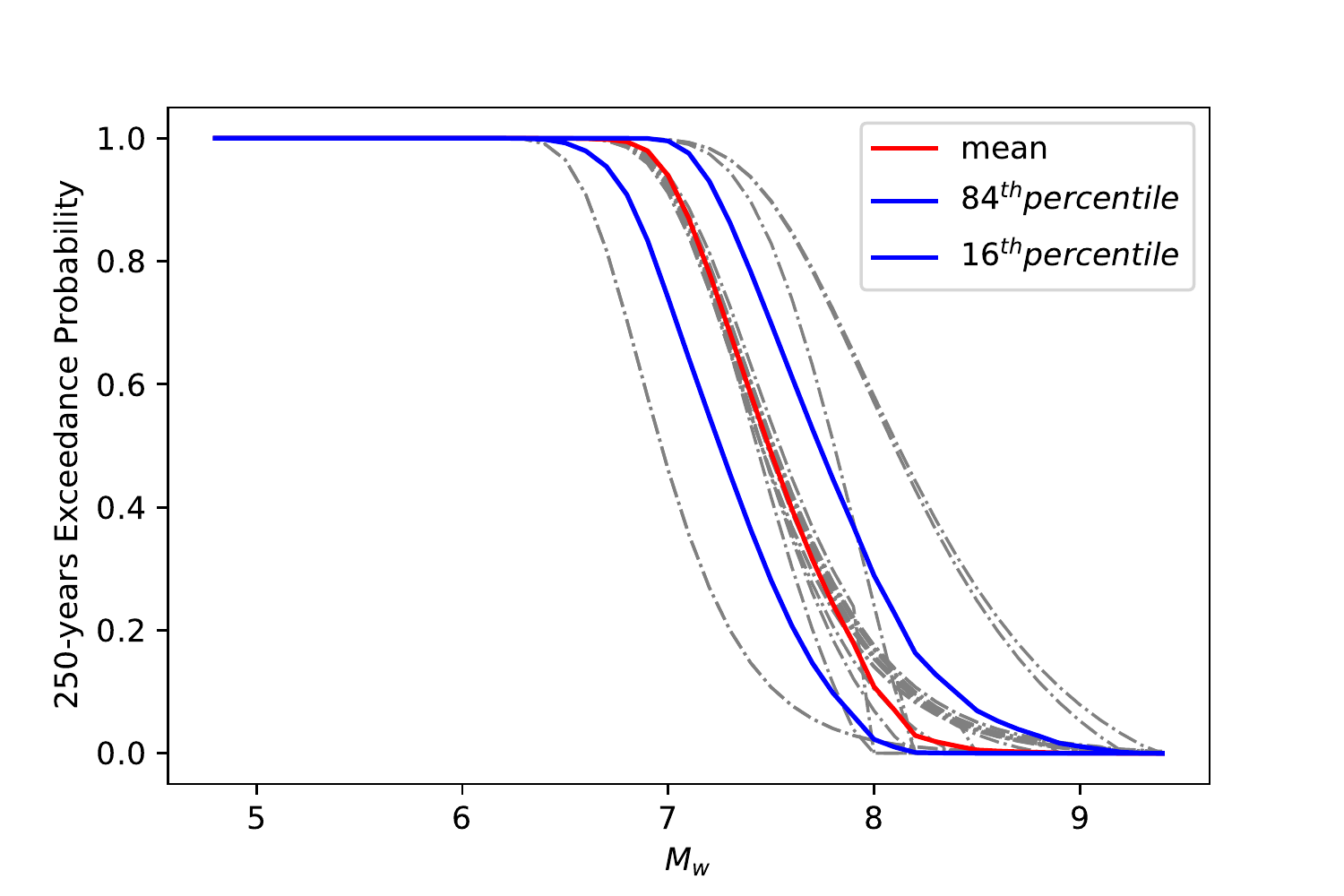}
	\includegraphics[width=7.8cm]{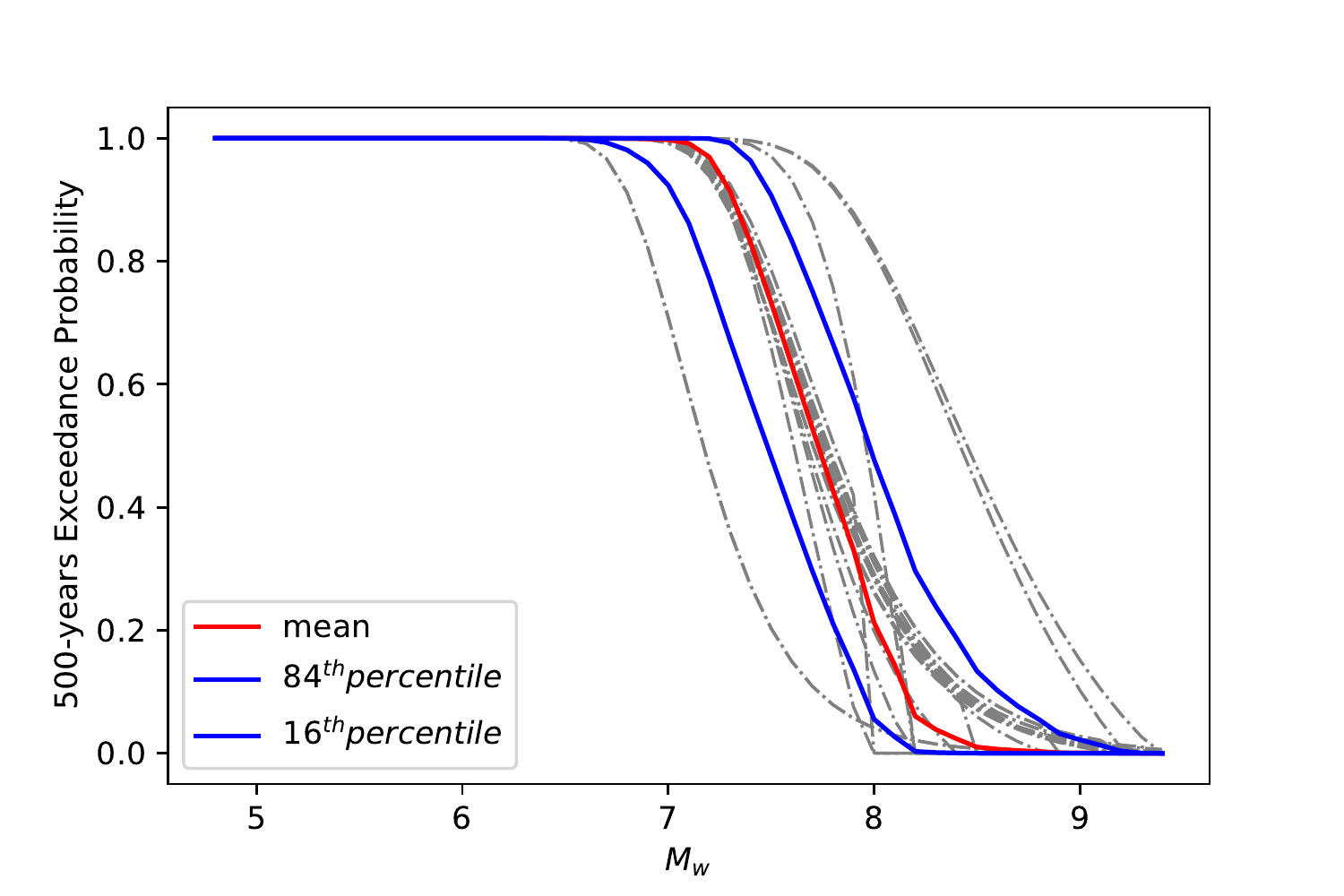}
	\includegraphics[width=7.8cm]{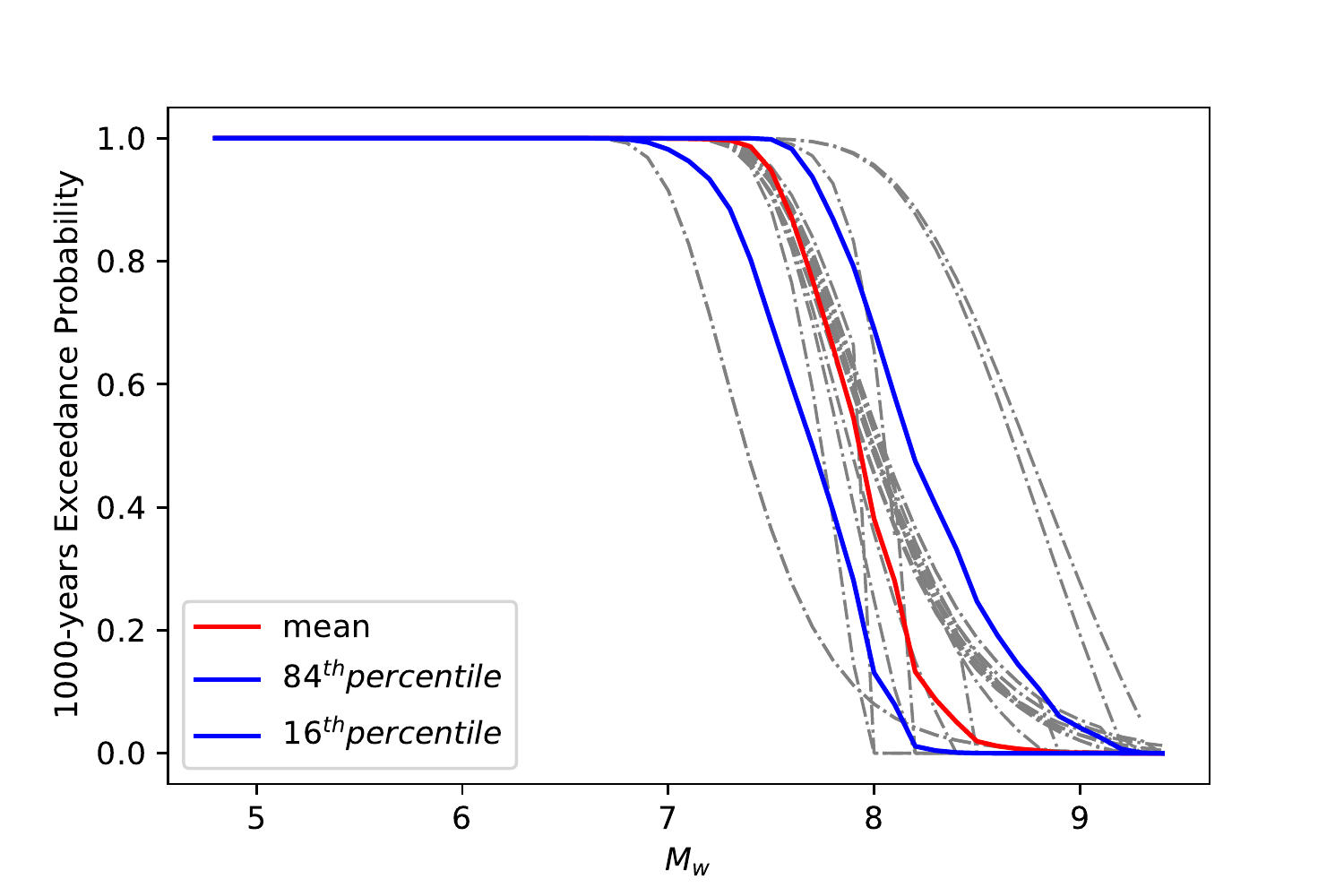}
	\caption{Earthquake probability of exceedance for our sample of $\Delta T$s: red and blue curves show the statistical description of ensemble model, i.e., mean and $16$th-$86$th percentiles, respectively. For comparison, all outcomes of the EV1 branches are also displayed in light gray.}
		\label{Fig:erthex}
\end{figure}

The earthquake probability of exceedance for the selected $\Delta T$s are depicted in Fig. \ref{Fig:erthex}. The ensemble model results from section~\ref{sec:ensemble} is shown through its statistical description, its mean, and the 16th-86th percentiles confidence intervals. For the sake of comparison,  the branches of EV1 ($\theta^{(E_j)}$) are also shown in light gray. In nearly all cases, the statistical description of the mean ensemble model is a good representative of EV1 branches Fig. \ref{Fig:erthex}. Henceforth, we use the value of mean ensemble for each magnitude to calculate tsunami probability exceedance curves and probability maps.
\subsection{Tsunami probability exceedance curves}\label{TPEC}
\begin{figure}[!ht]
	\centering
	\includegraphics[width=7.8cm]{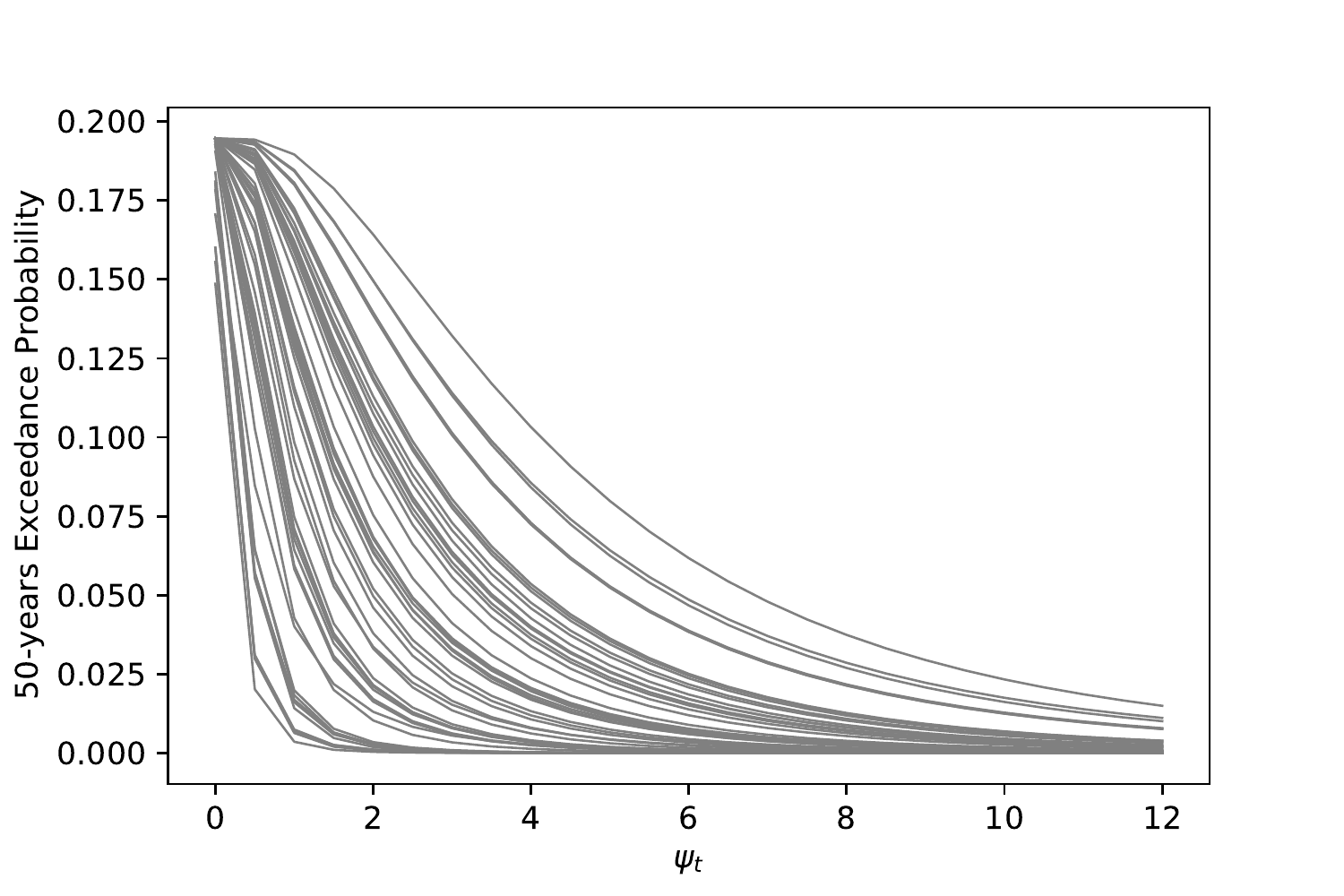}
	\includegraphics[width=7.8cm]{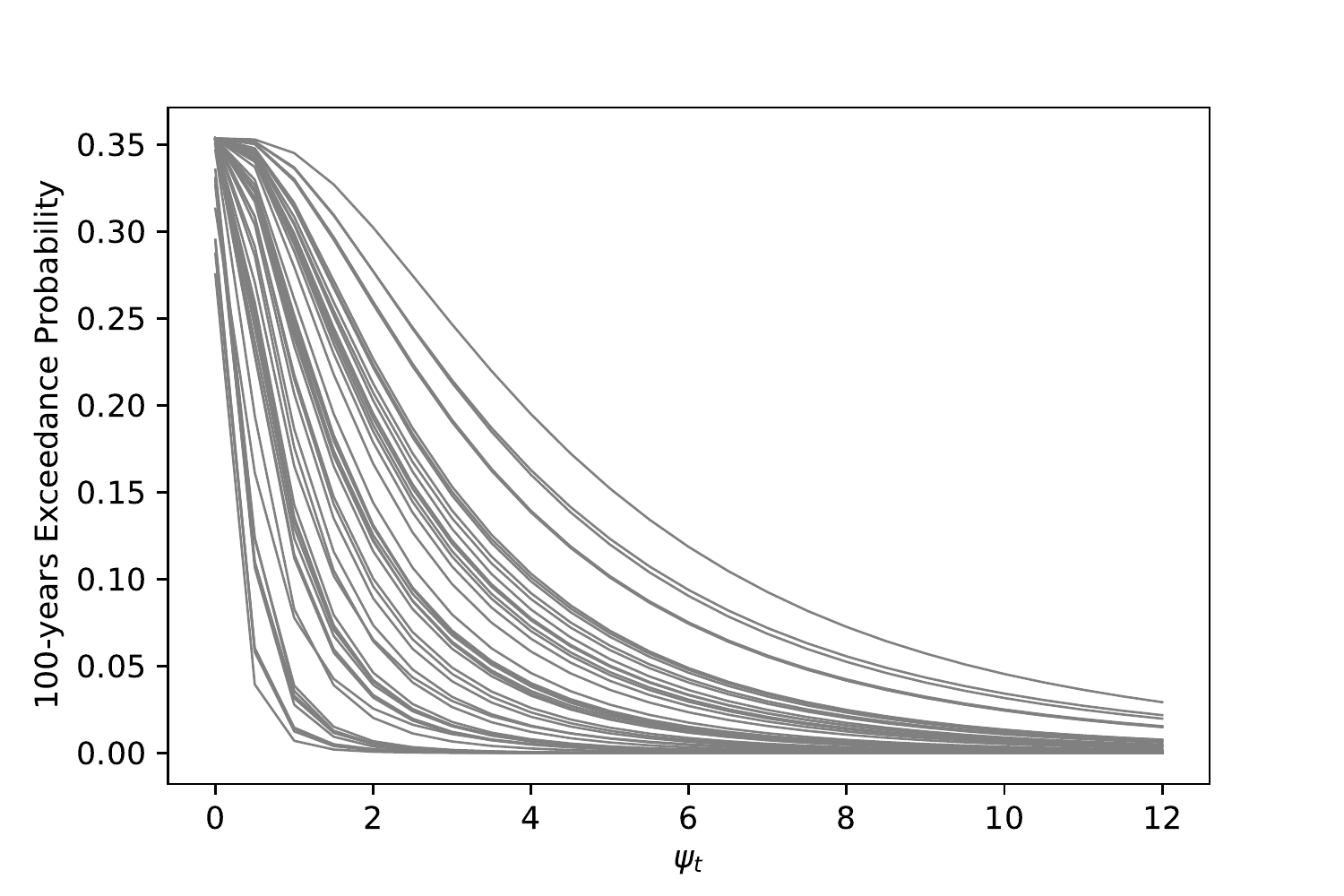}
	\includegraphics[width=7.8cm]{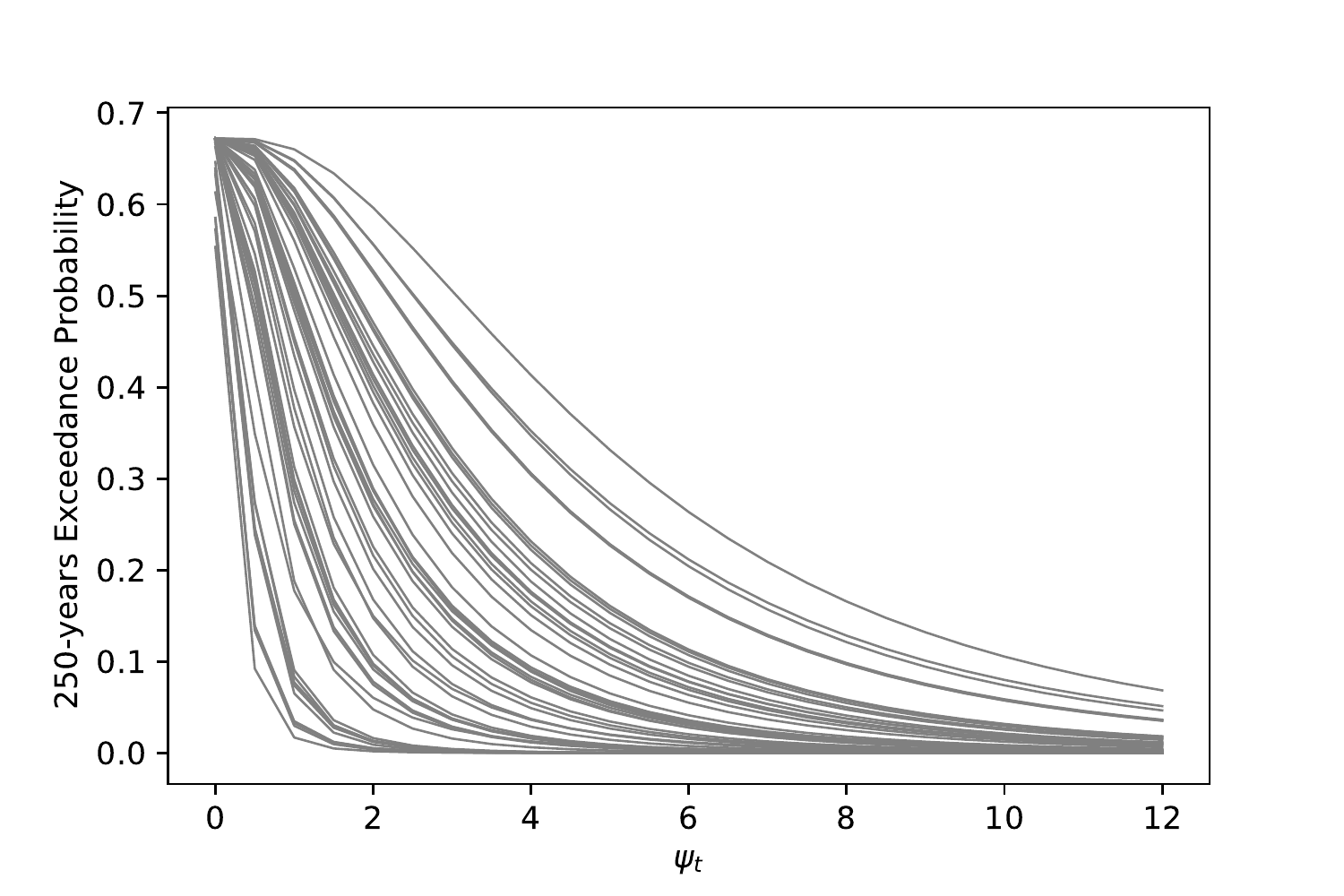}
	\includegraphics[width=7.8cm]{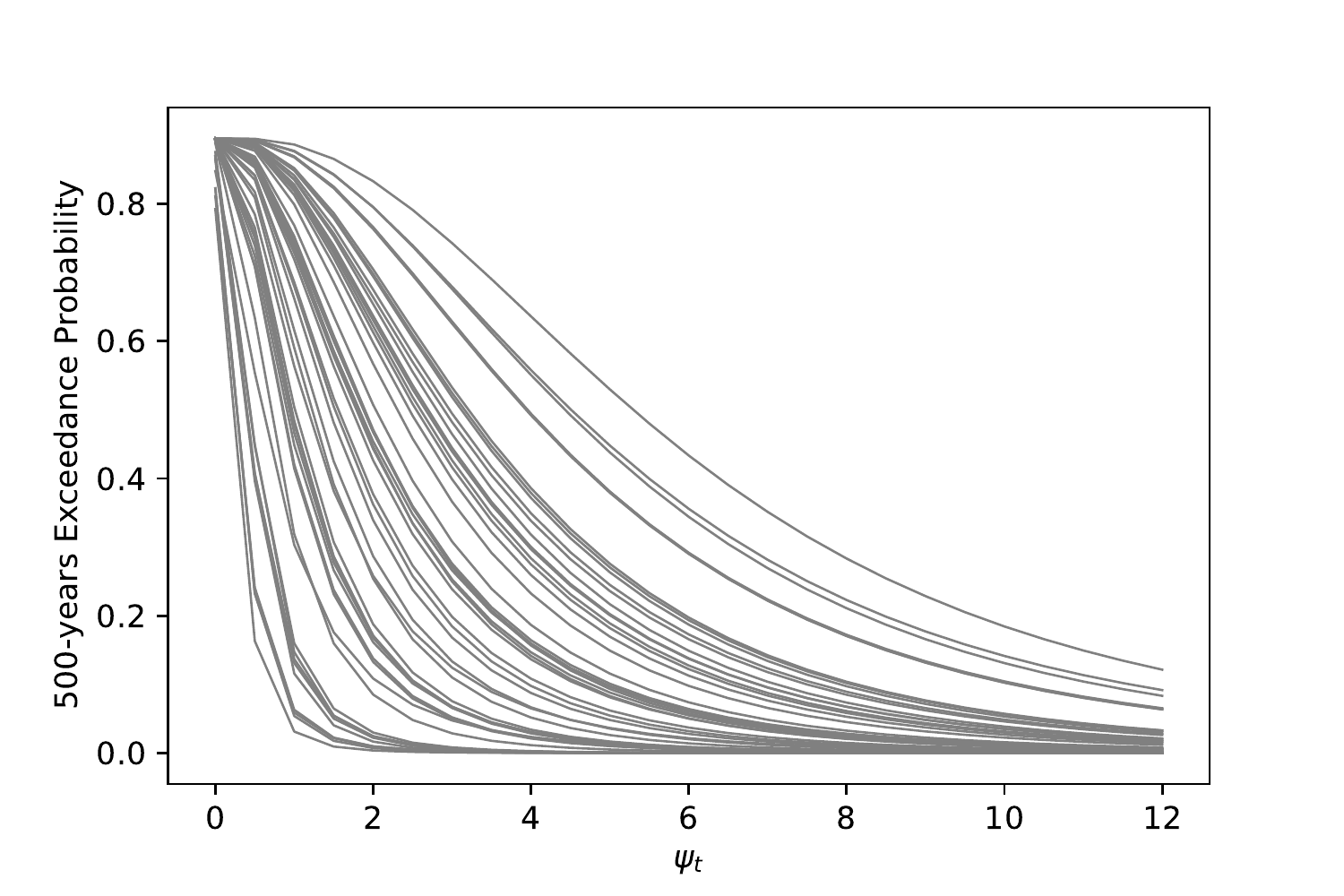}
	\includegraphics[width=7.8cm]{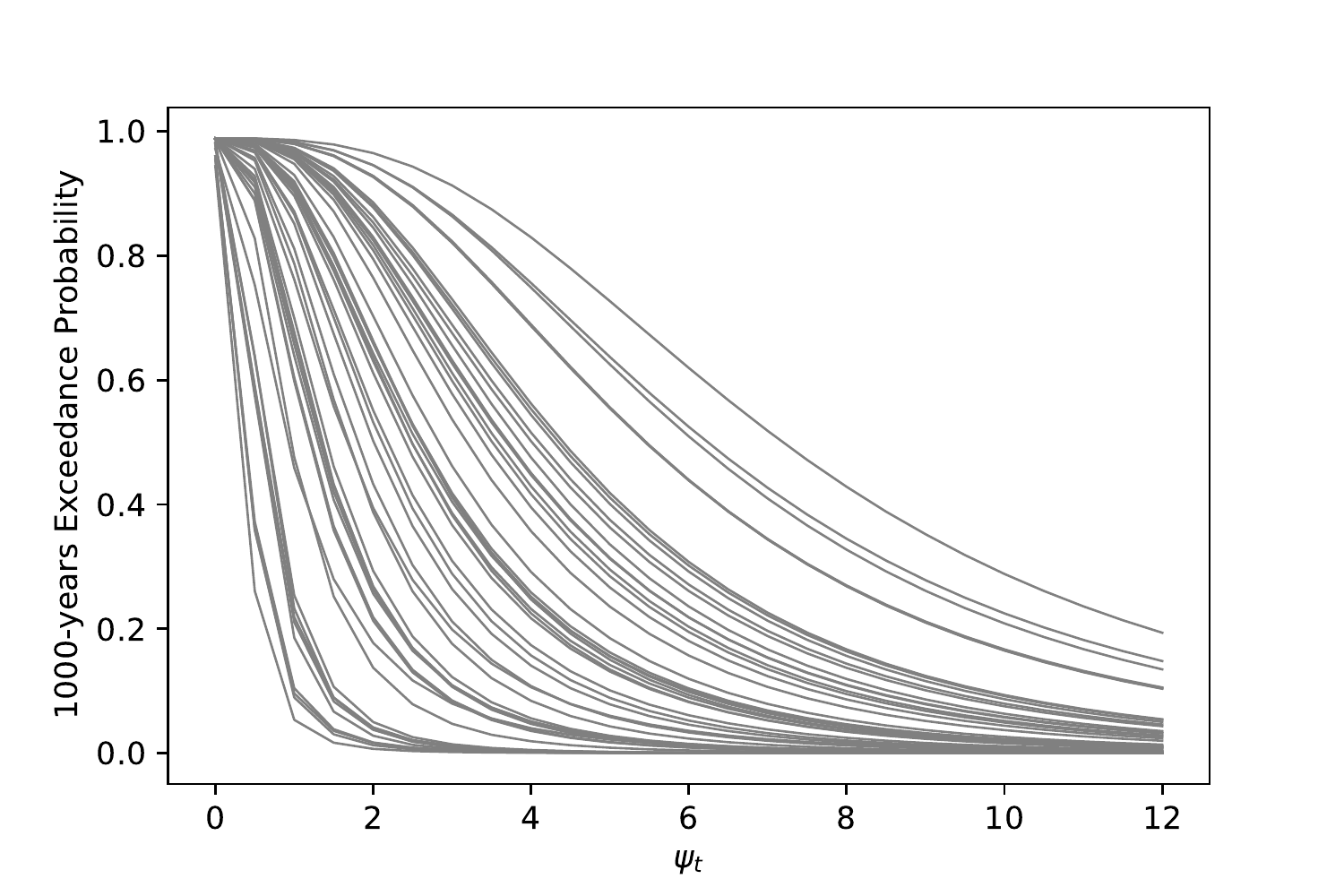}
	\caption{Tsunami probability of exceedance for a sample of $\Delta T$s at different PoIs along the Iran and Pakistan coasts.}
		\label{Fig:tsuex}
\end{figure}

\begin{figure}[!ht]
	\centering
	\includegraphics[width=7.8cm]{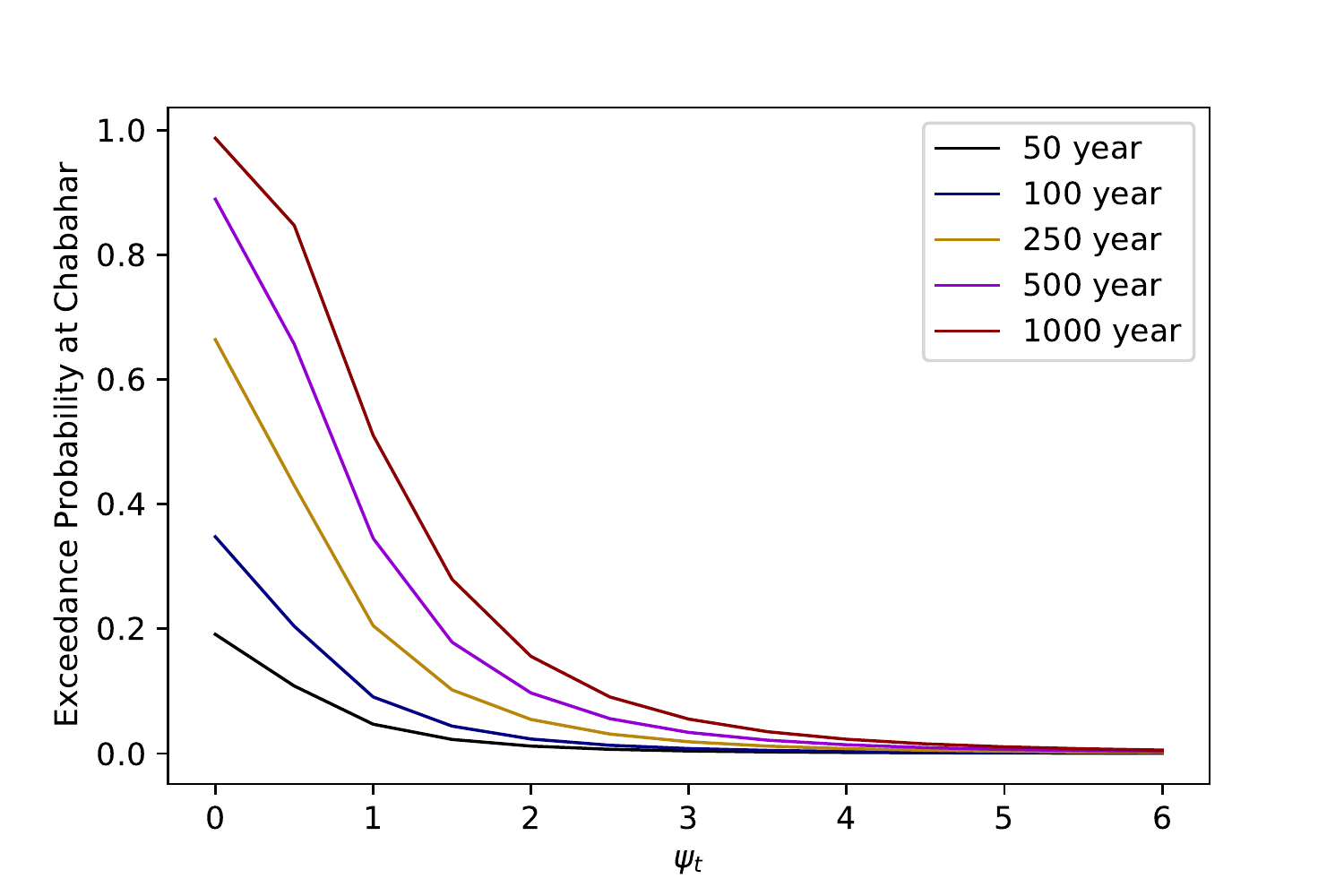}
	\includegraphics[width=7.8cm]{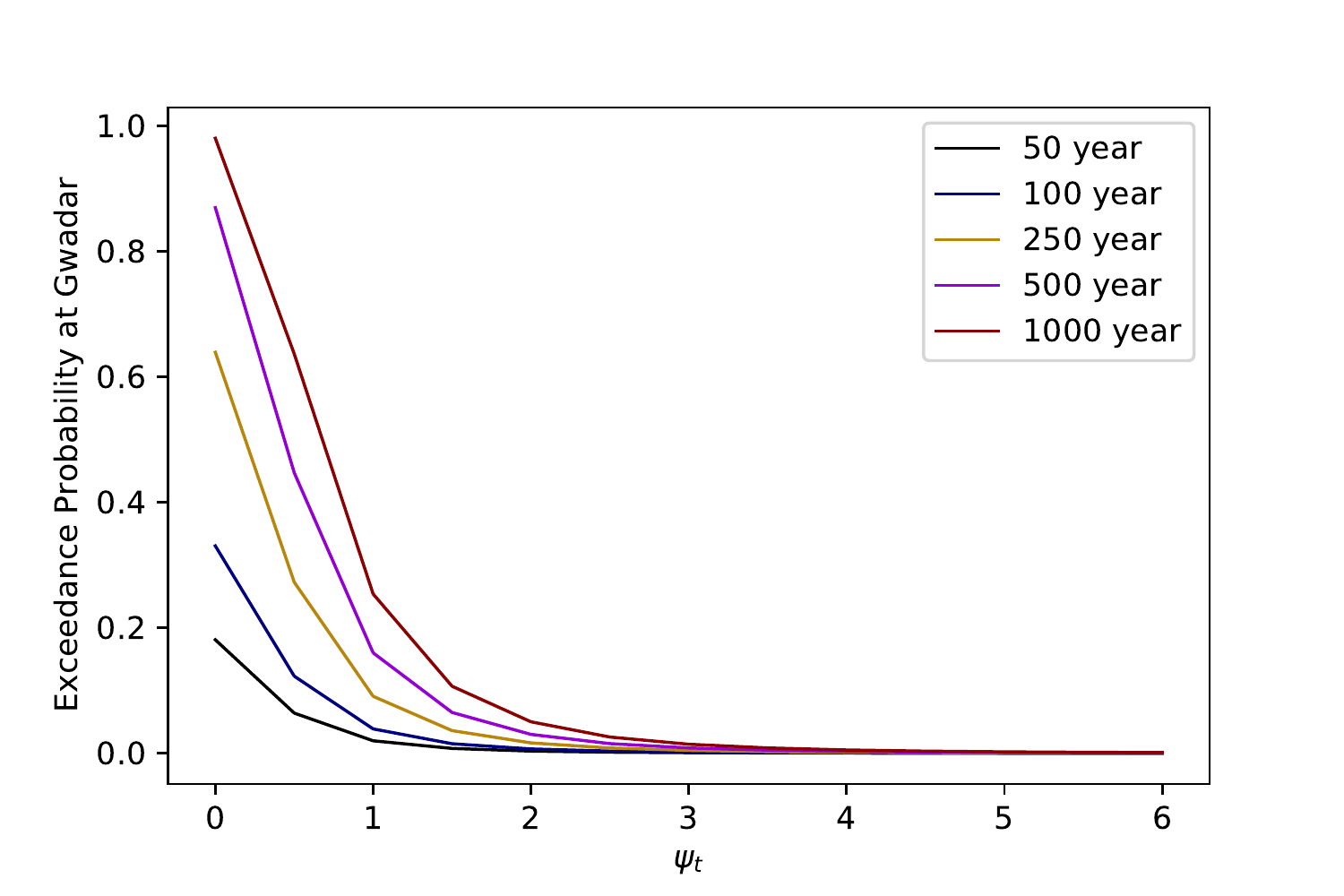}
	\includegraphics[width=7.8cm]{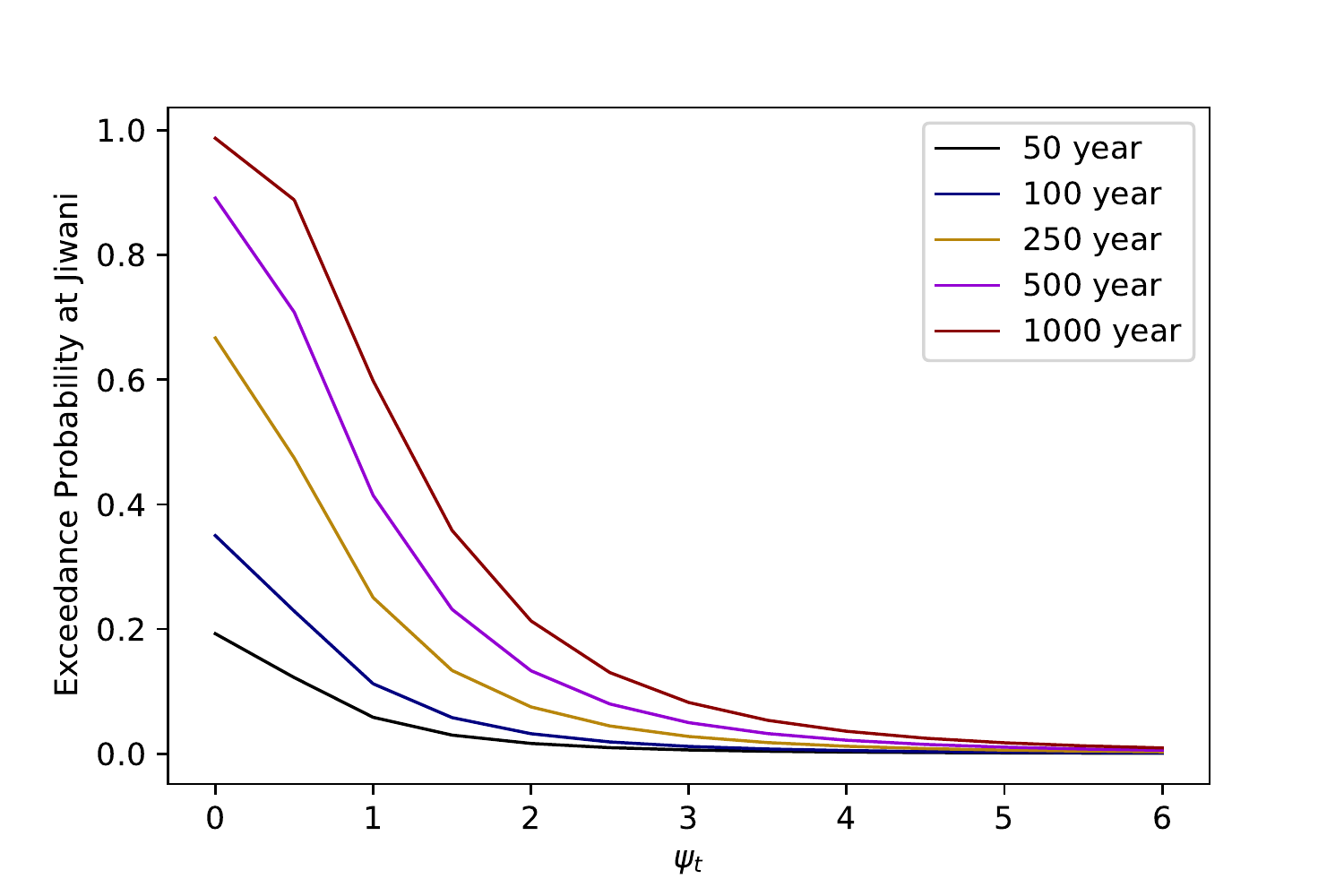}
	\includegraphics[width=7.8cm]{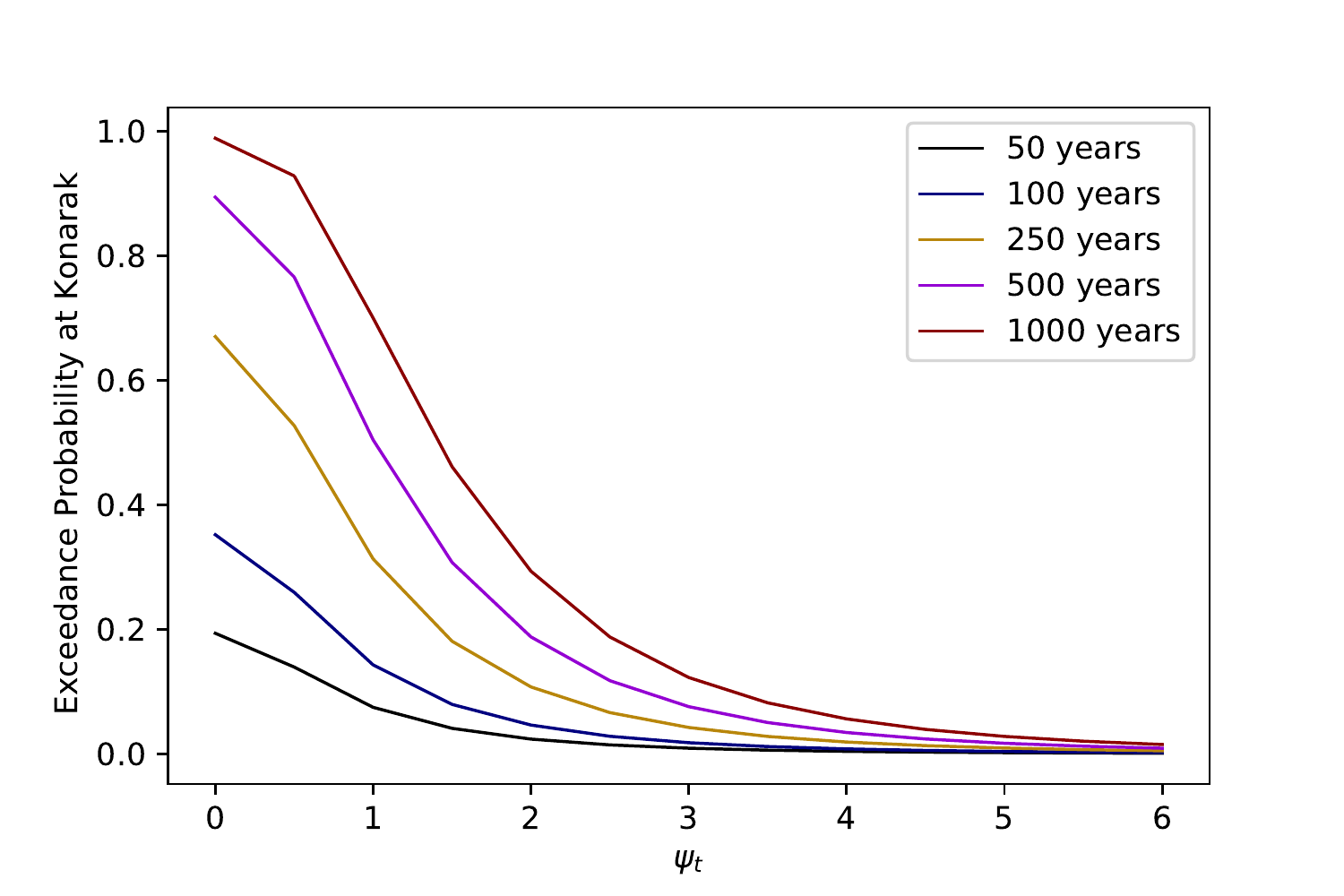}
	\includegraphics[width=7.8cm]{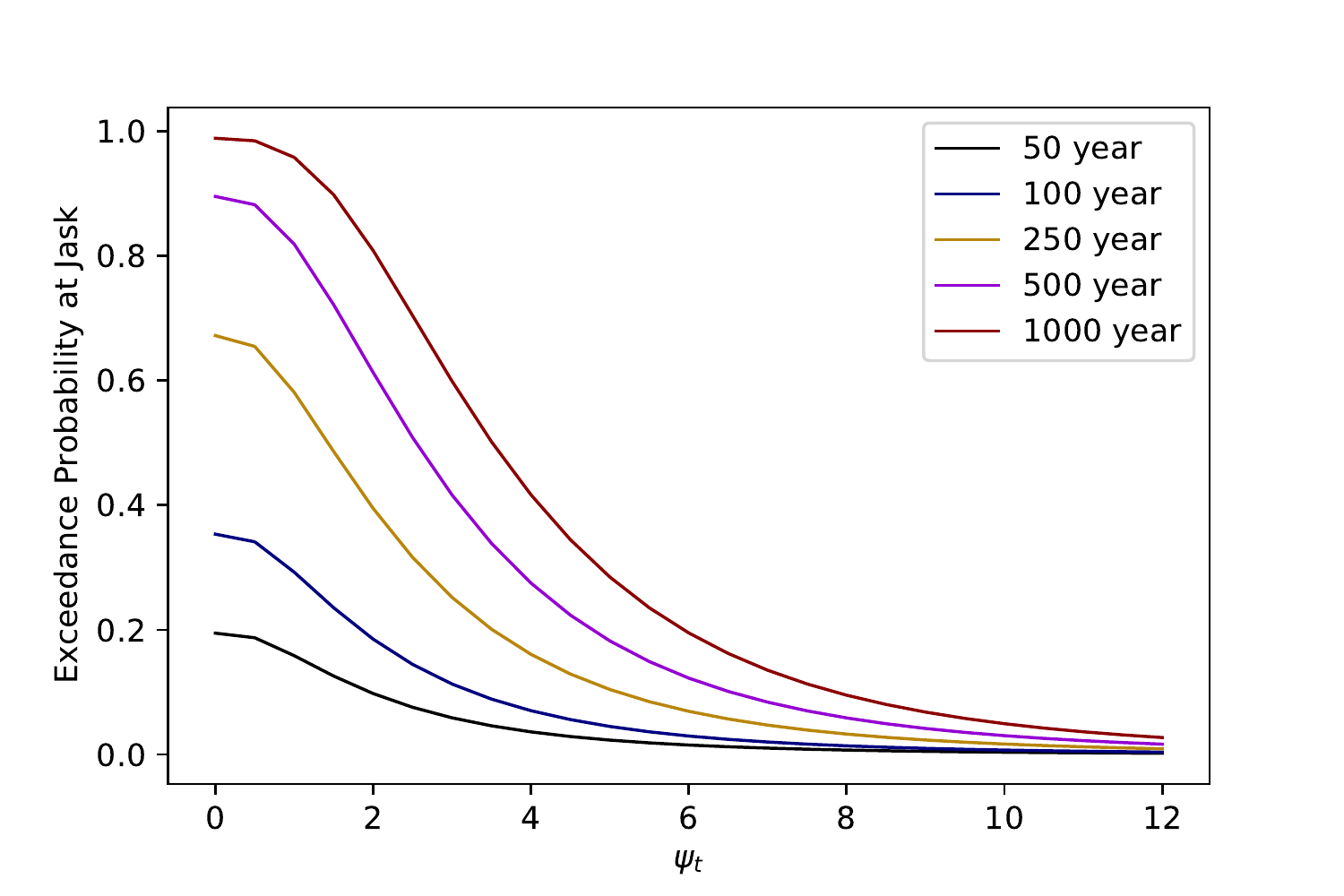}
	\includegraphics[width=7.8cm]{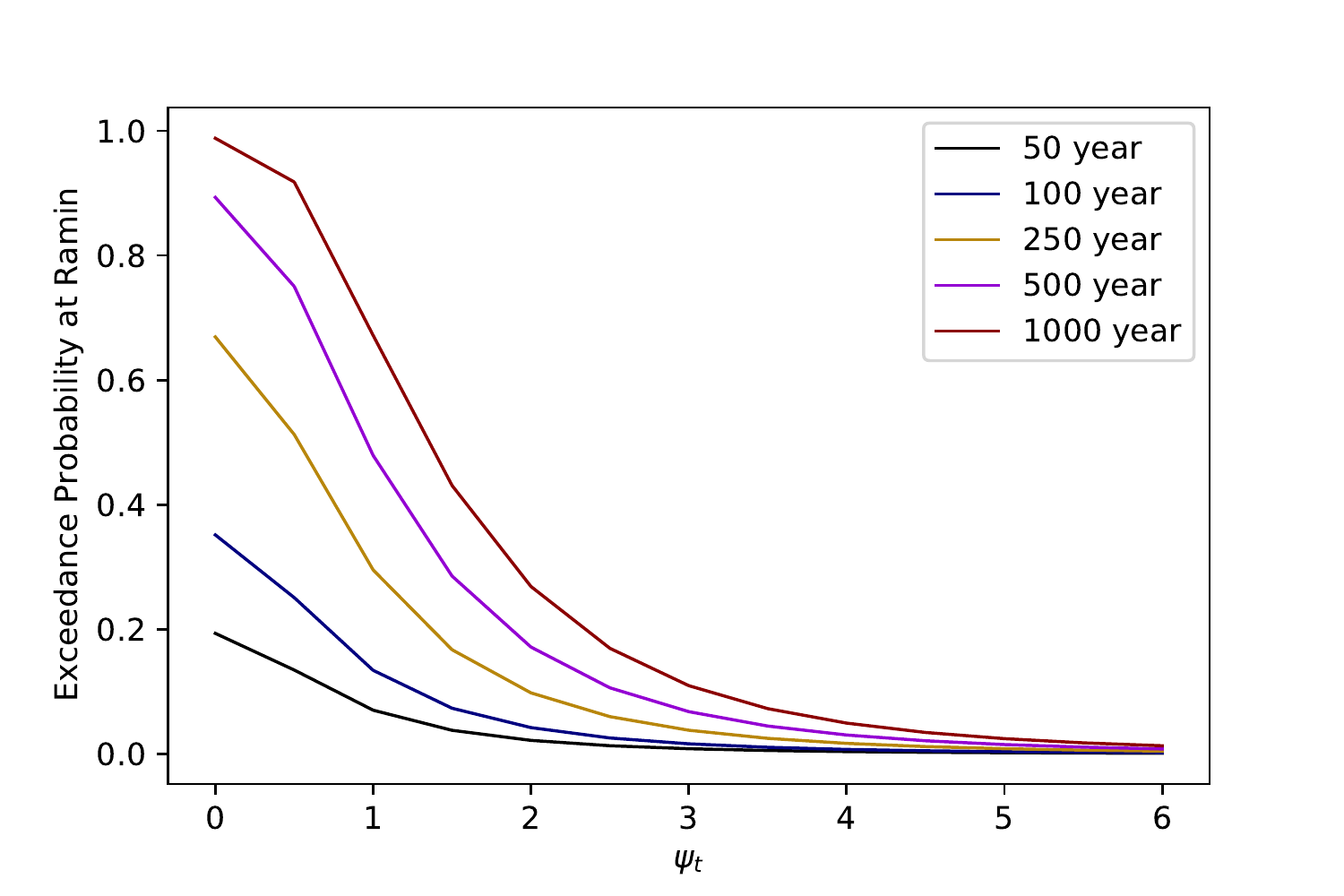}
	\caption{Tsunami probability of the selected $\Delta T$s exceedance for the selected six PoIs near main cities.}
		\label{Fig:RPC}
\end{figure}

Using the equations described in section~\ref{Sec:calculation}, we calculated  the probability of exceedance from a set of tsunami height thresholds $\psi_t=\{0.5,1,1.5,..,11.5,12\}$ m and generated hazard curves at different PoIs incorporating all uncertainties described in the previous sections. The results are given in Fig.~\ref{Fig:tsuex}. The hazard curves for each PoI are shown in gray. The results show that  the spread of hazard curves for different locations of the Makran coast is remarkably large. As an example, $P^{\text{tot}}( \psi > 3 ,\Delta T= 50)$ ranges from $0$ to $13.5\%$ for different PoIs. By increasing $\Delta T$, this range opens up and it reaches $25\%, 52\%, 74\%$, and $91\%$ for the return periods of 100, 250, 500, and 1000 years, respectively. It is thus not wise to consider a mean (or percentile) of PoI hazard curves for any purpose in the coasts of Iran and Pakistan. Hence, we selected six main PoIs close to the major cities of the Makran region, namely, Chabahar, Konarak, Jask, Ramin, Jiwani, and Gwadar, to explore the results in detail.
Fig.~\ref{Fig:RPC} shows the tsunami probability exceedance curve at the above mentioned six major cities for different return periods. For a 50-year return period, the probability of exceedance does not vary much among different cities. However, this difference becomes significant with increasing $\Delta T$. For instance, for $P^{\text{tot}}( \psi > 1 ,\Delta T= 1000)$, it ranges from $26\%$ in the west (Gwadar) to $94\%$ in the east (Jask). Moreover, the probability that tsunami height exceeds 4 m is low (less than $10\%$) near all major cities except for Jask, with the probability of $53\%$.

\subsection{Sensitivity analysis}

\begin{figure}[!ht]
	\centering
	\includegraphics[width=7.8cm]{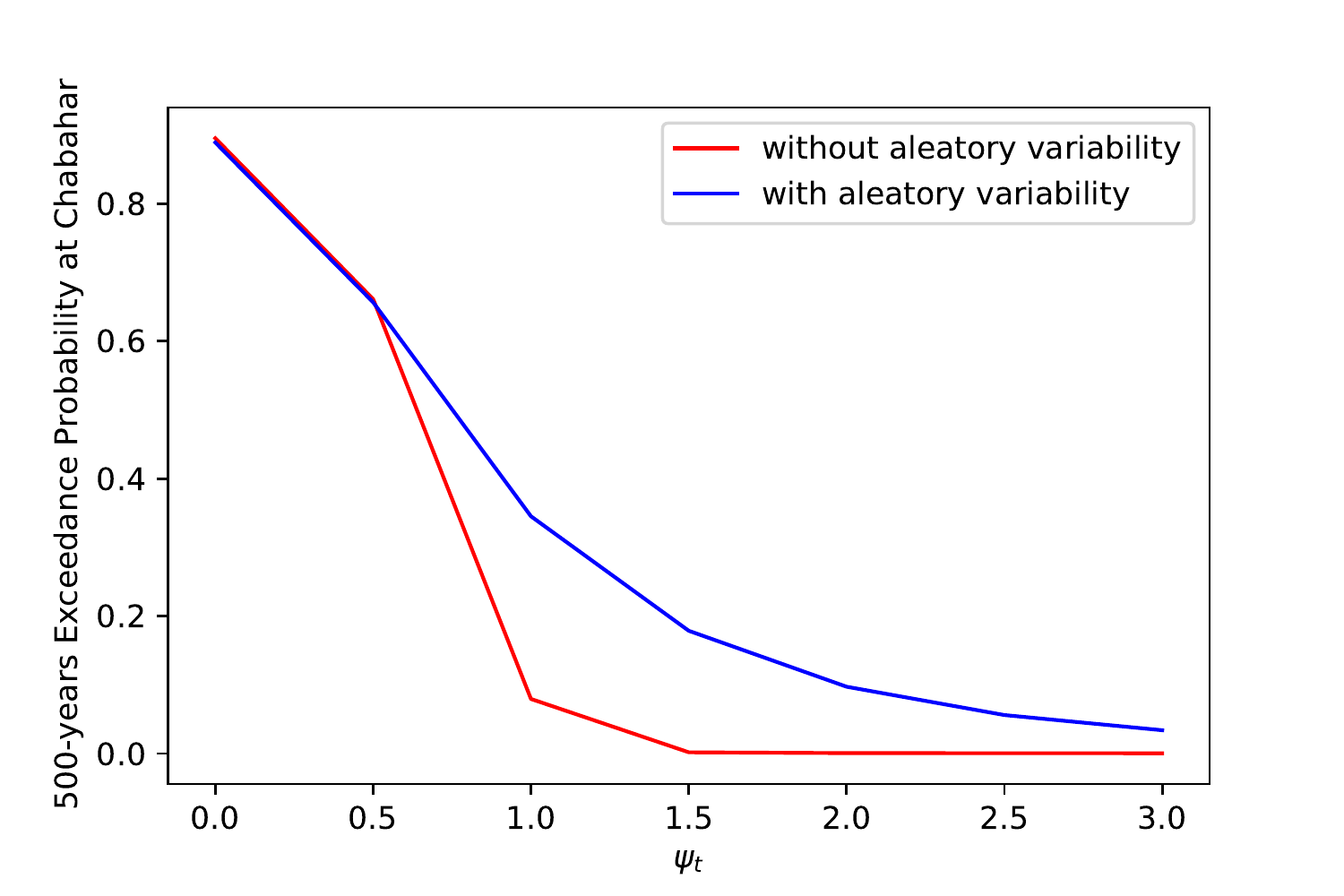}
	\includegraphics[width=7.8cm]{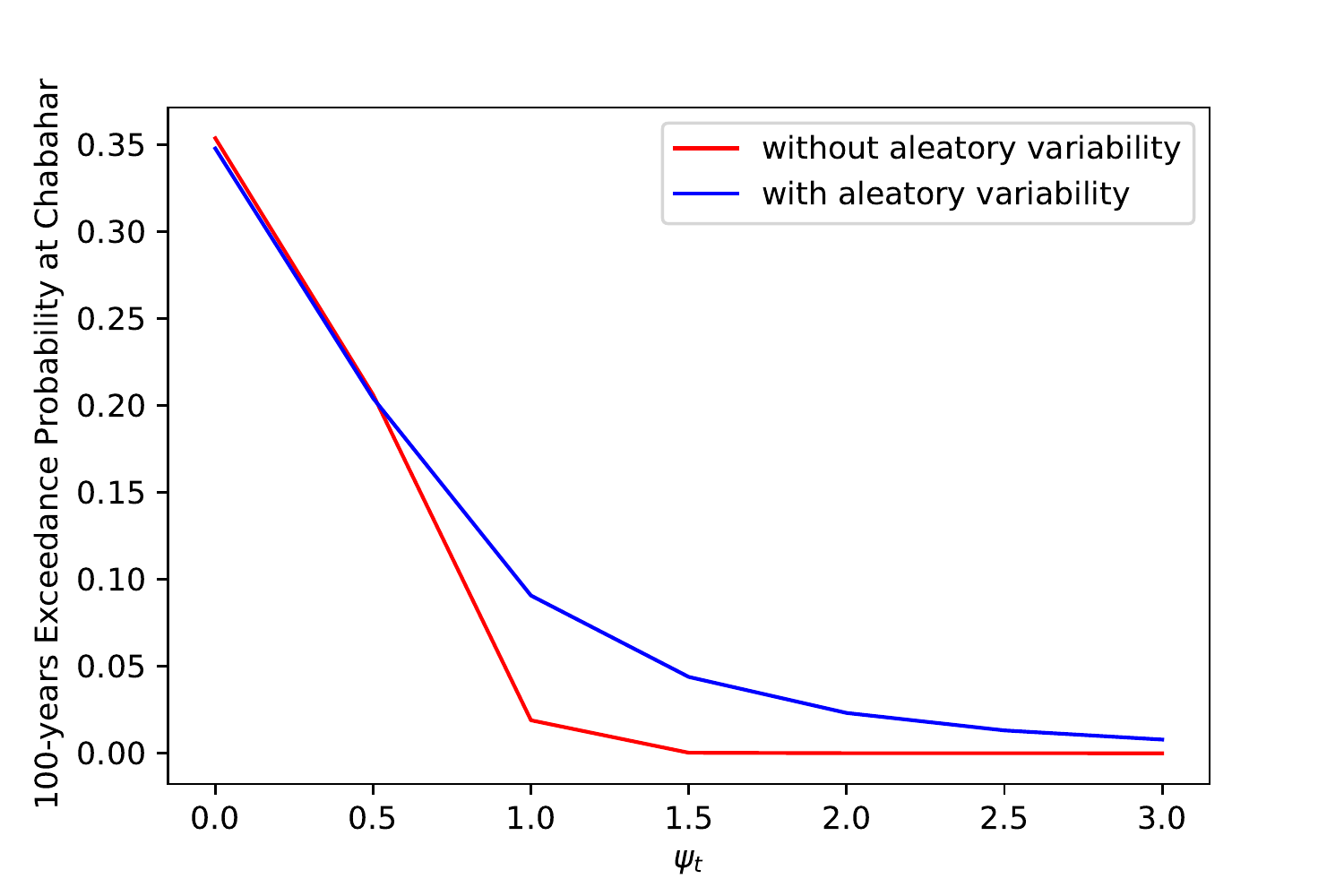}
	\caption*{(a)}
	\includegraphics[width=11cm]{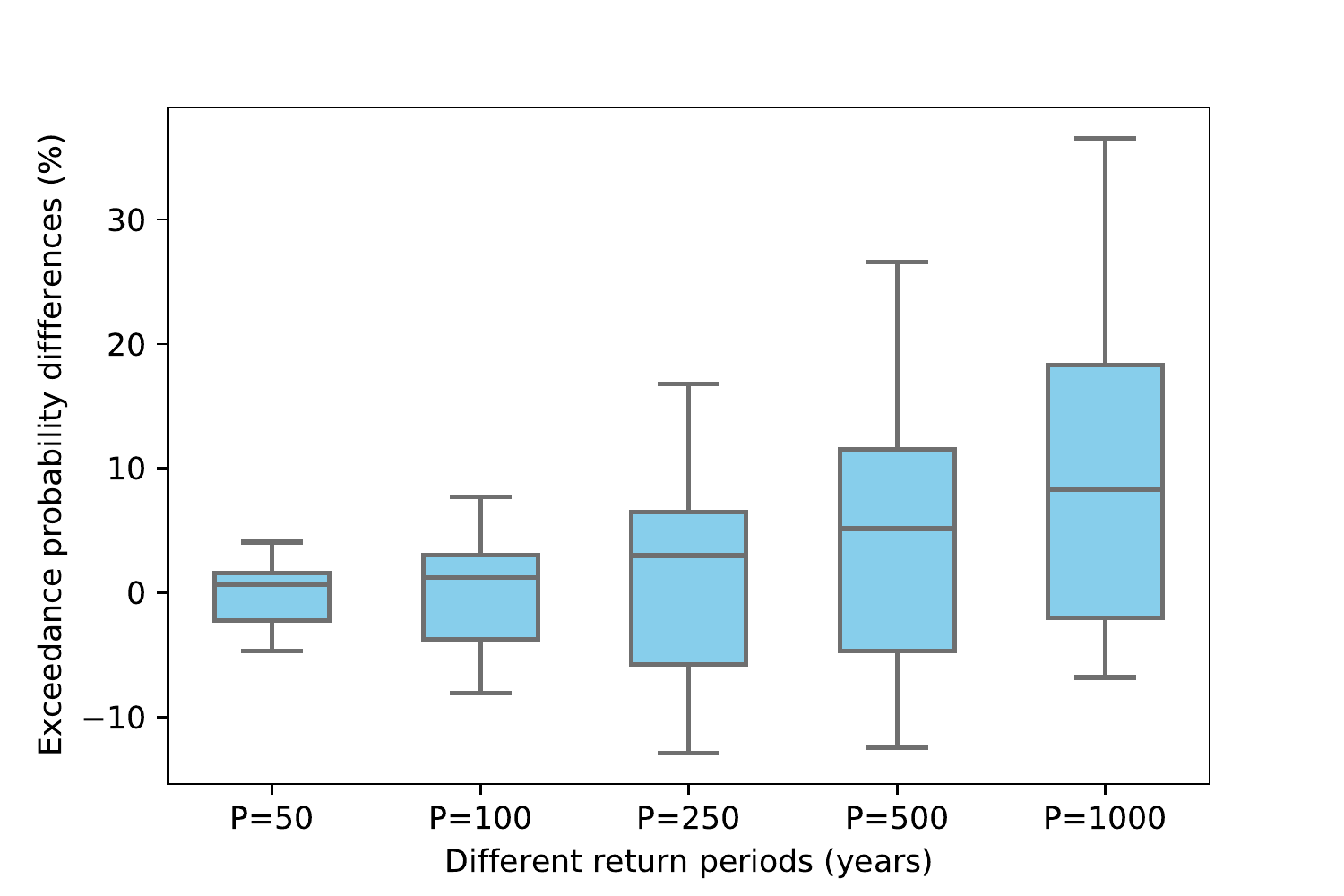}
	\caption*{(b)}
	\caption{(a) Exceedance curve of Chabahar as a random PoI for 100- and 500-year return periods; blue and red curves show the probability of exceedance in the presence and absence of the aleatory variability, respectively; (b) box plot showing the differences in exceedance probability ($\%$) for different $\Delta T$s with and without the presence of the aleatory variability for all  PoIs.
		\label{Fig:dif}}
\end{figure}

The effect of inclusion of the aleatory variability introduced in section~\ref{Sec:aleatory} is shown in Fig.~\ref{Fig:dif}. Fig.~\ref{Fig:dif} (a) illustrates the probability of exceedance in the presence and absence of the aleatory variability  at one random  PoI (i.e., Chabahar) for two return periods (100 and 500 years). The inclusion of the aleatory variability has a significant effect on the probability of exceedance, which increases for a longer return period. As an example, the differences between  $P^{\text{tot}}( \psi > 1 ,\Delta T, \text{Chabahar})$ with and without the aleatory variability are $8\%$ and $26\%$ for $\Delta T= 100$ and $500$ years, respectively. To obtain a better interpretation, we also calculated this difference for all PoIs and $\Delta T$s, see Fig.~\ref{Fig:dif} (b). In summary, omitting the aleatory variability mostly leads to a noticeable underestimation with a median of $10\%$ for all  PoIs, reaching  and it even reaches $40\%$ at somewhere for 1000-year return period.

\subsection{Probability maps}
\begin{figure}[!htp]
	\centering
	\includegraphics[width=7.8cm]{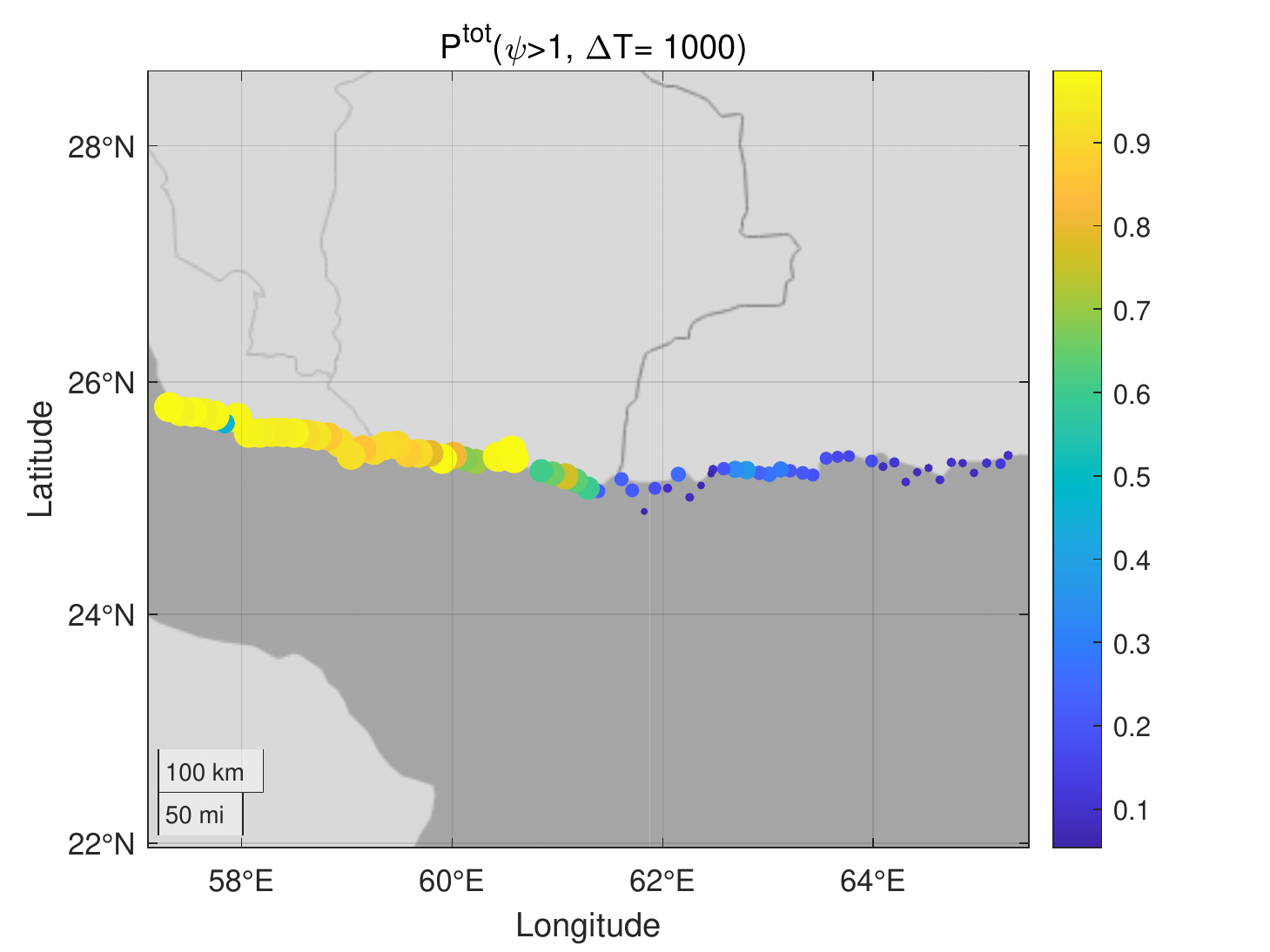}
	\includegraphics[width=7.8cm]{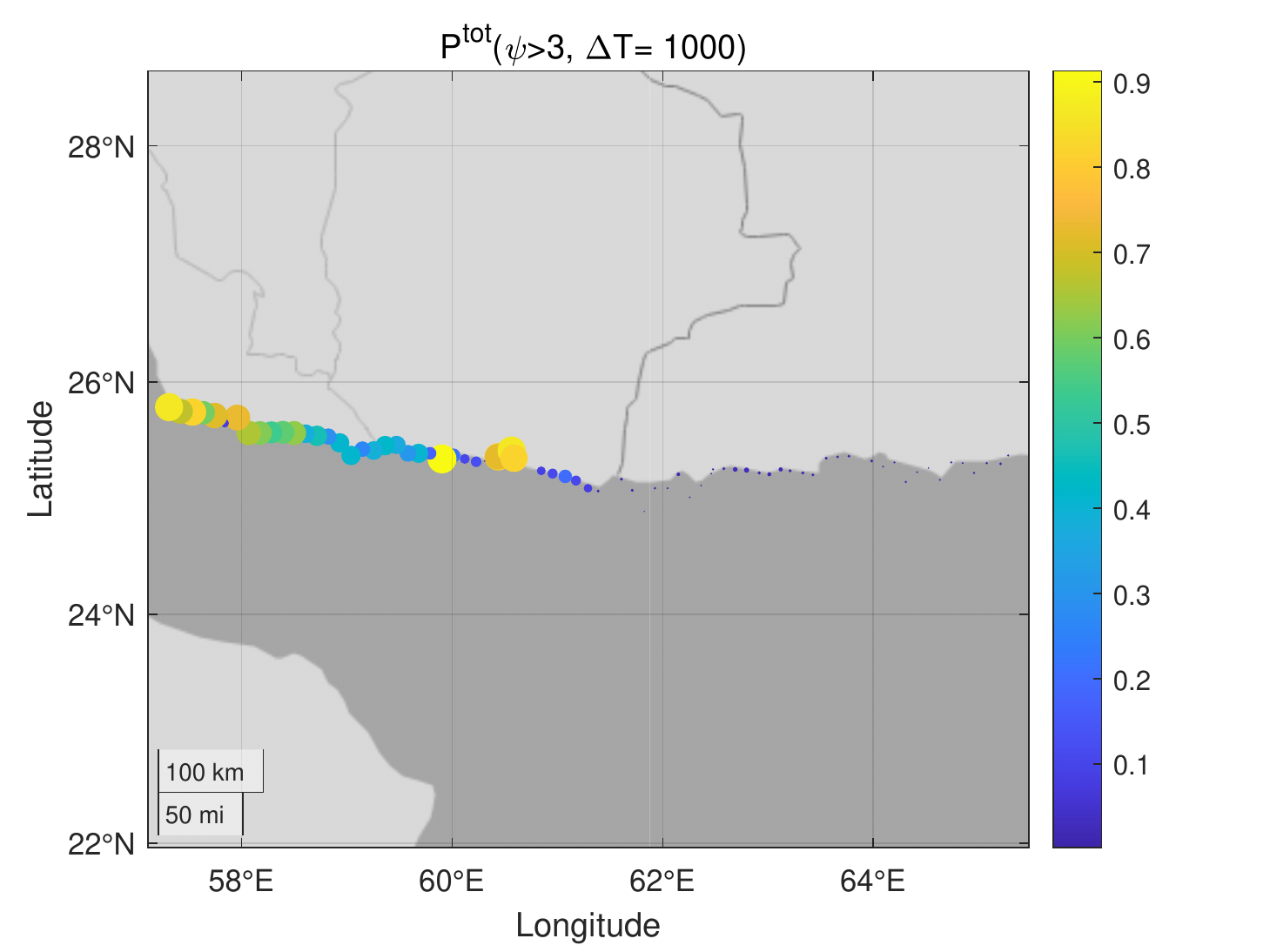}	\includegraphics[width=7.8cm]{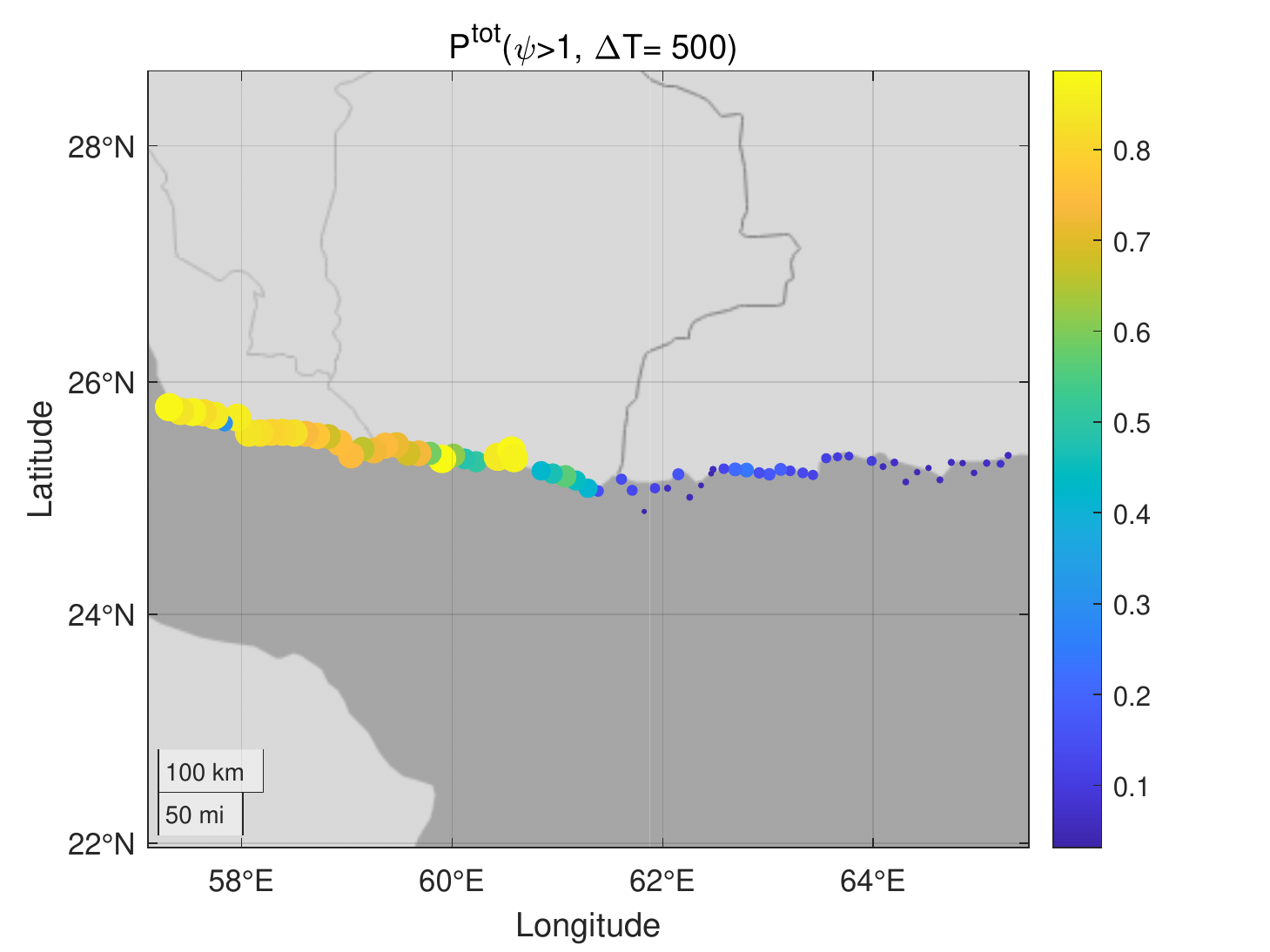}
	\includegraphics[width=7.8cm]{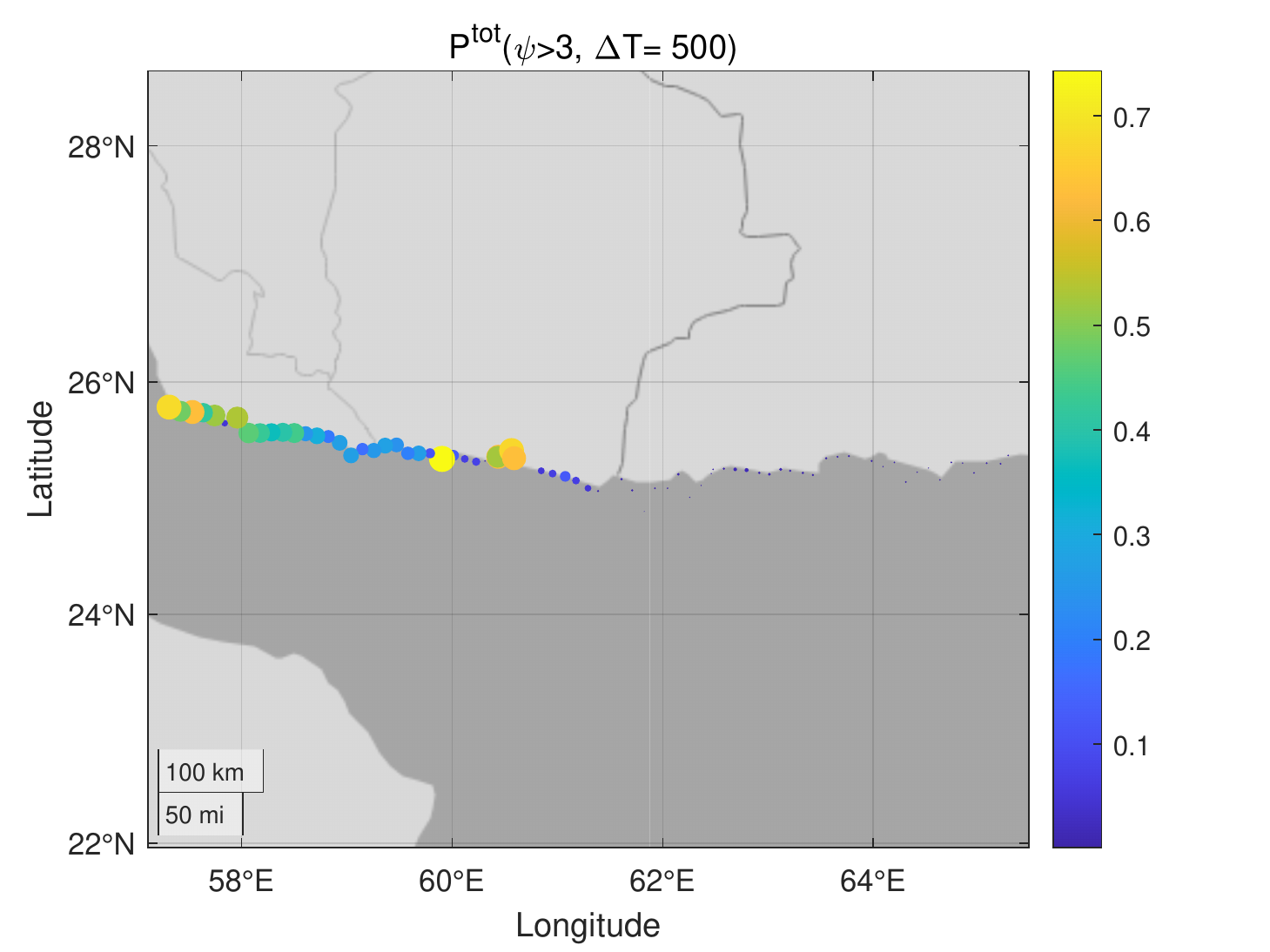}
	\includegraphics[width=7.8cm]{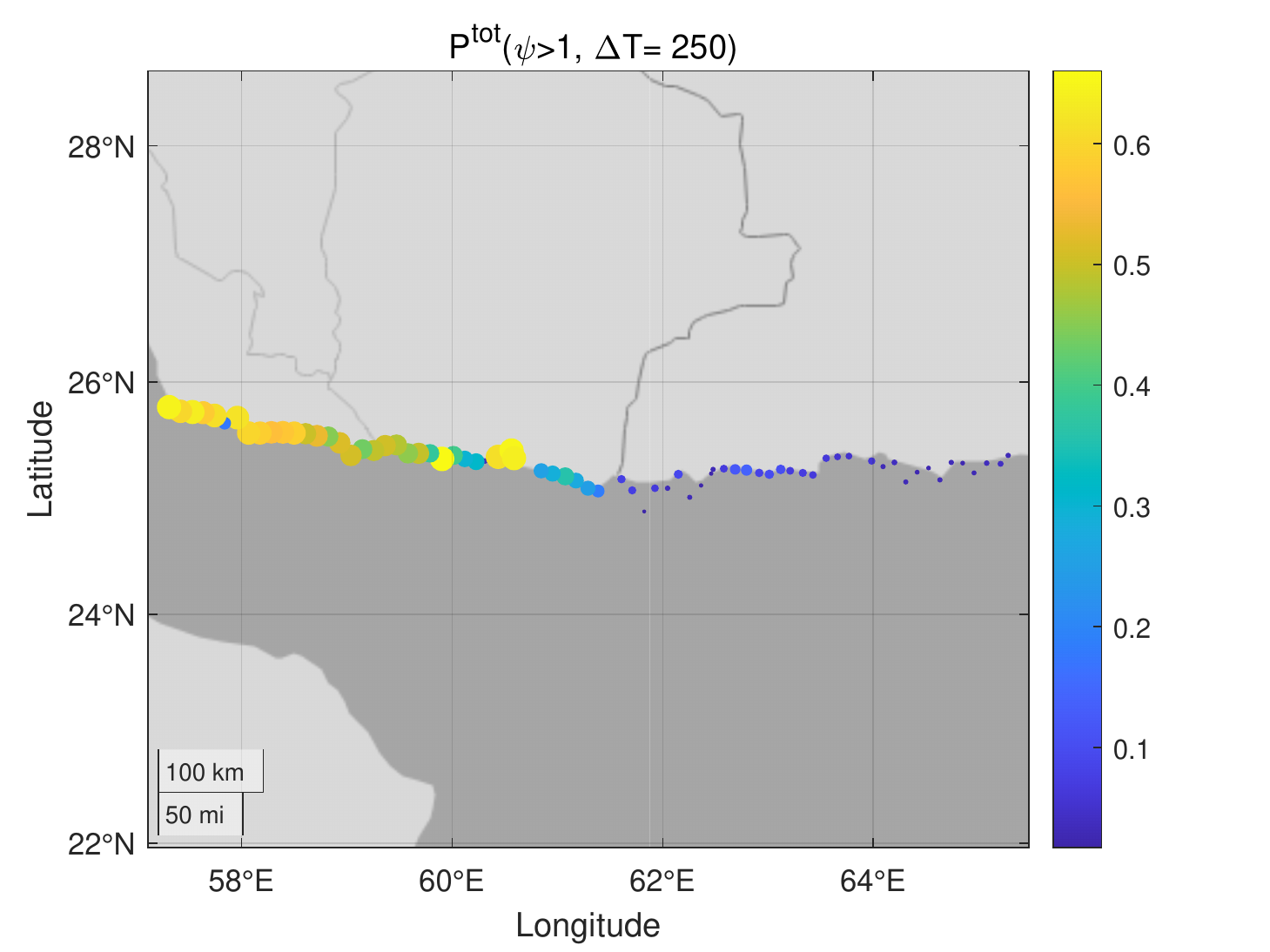}
	\includegraphics[width=7.8cm]{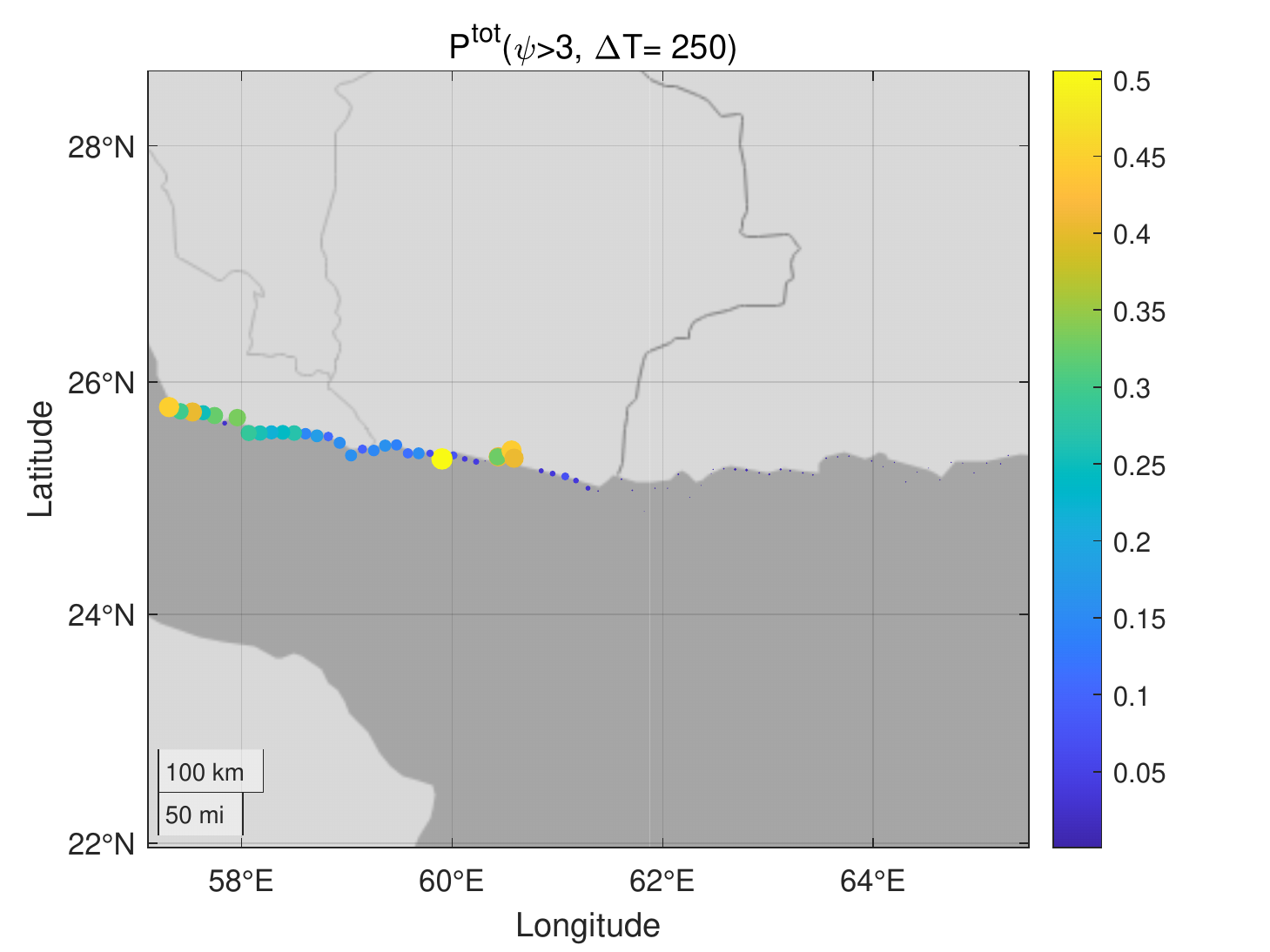}
	\includegraphics[width=7.8cm]{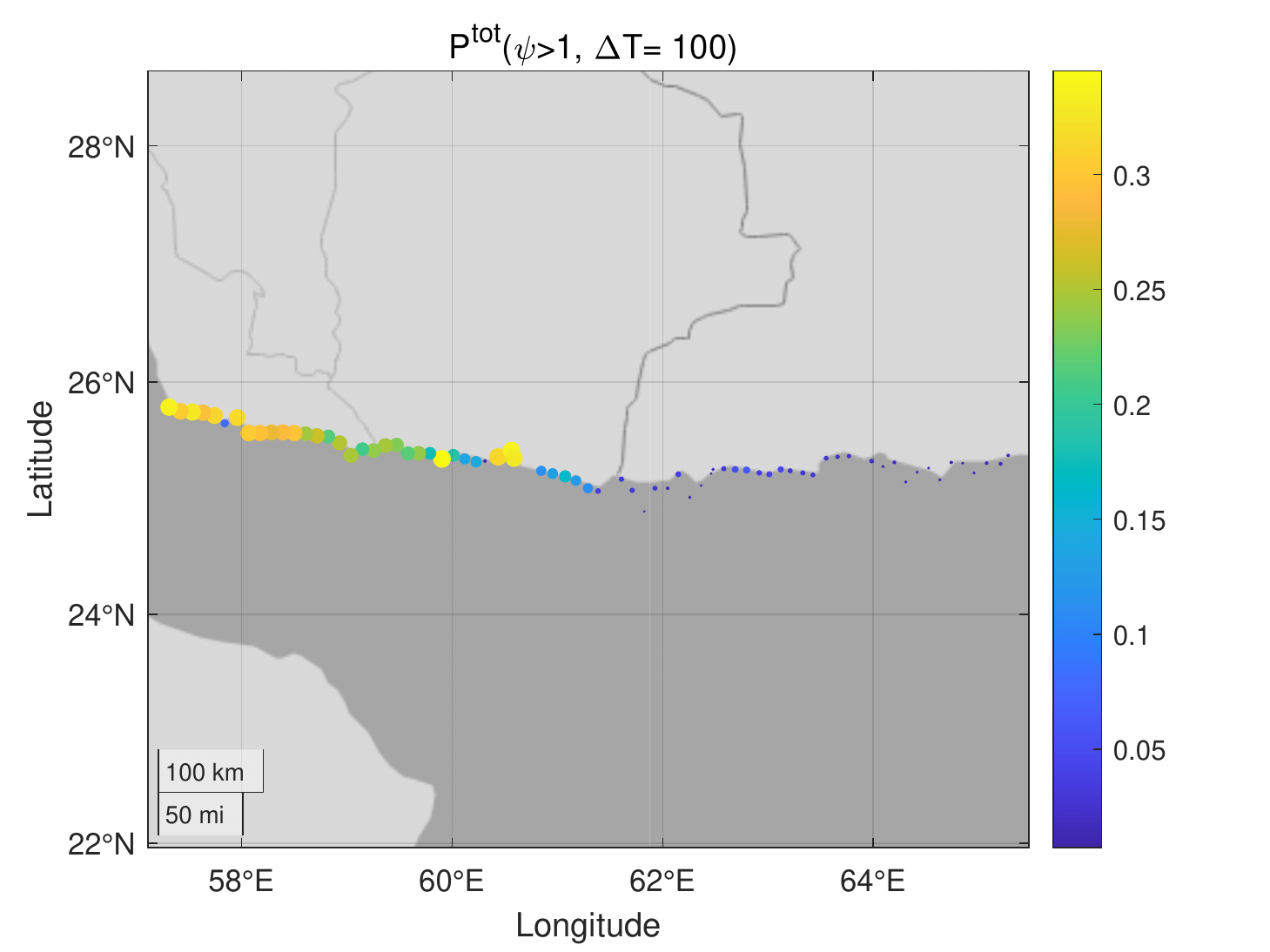}
	\includegraphics[width=7.8cm]{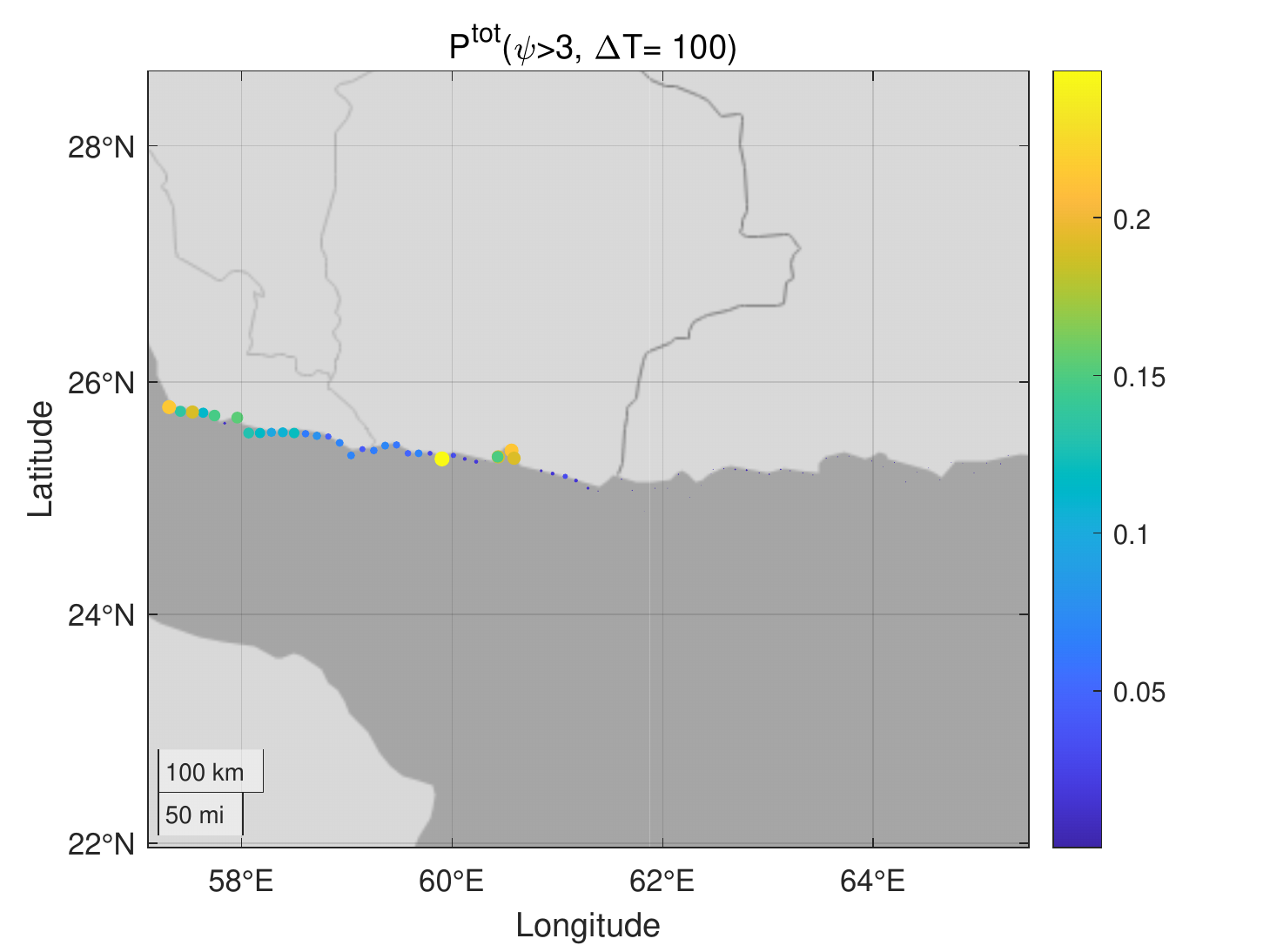}
	\caption{Maps of tsunami probability exceeding 1 and 3 m for different $\Delta T$s along the entire coast of Iran and Pakistan.}
		\label{Fig:pmap}
\end{figure}

We used a probability map to assess hazard along the entire coast irrespective of population density, which is crucial for prioritizing tsunami mitigation plans and city development in low-population areas. Most literature and mitigation plans focus on specific populated areas (e.g., \citep{akbarpour2017tsunami, payande2015tsunami}. Fig.~\ref{Fig:pmap} illustrates tsunami probability maps exceeding from two selective thresholds, $\psi_t=1,3$ m with different return periods. The probability of exceedance is much more intense in the west. Furthermore, in some rural areas (e.g., Tis and Tang) neighboring Chabahar, the probability that tsunami height will exceed 3 m for return periods of 100 and 1000 years is approximately $25\%$ and $89\%$, respectively. Notably, this is almost $6$ to $7$ times higher than that in Chabahar. Owing to the small distances between these regions, the inundated area at Chabahar may be affected. Inundation maps are beyond the scope of this study, and we plan to address them in a future work.

\section{Conclusions}
The MSZ is one of the two sources of tsunamis in the Indian Ocean, and has the potential of generating large tsunamis that threaten neighboring countries of Iran, Oman, and Pakistan. However, a fortune lack of large tsunamis recently has led to a false sense of safety between community leaders and residents, which may negatively affect the area’s vulnerability and resilience against future tsunamis. In addition, a short historical record compared to return period for major subduction zones makes it difficult to conclude the potential risks of future tsunamis. Therefore, in this study, we assessed the potential seismogenic zone, maximum magnitude, and recurrence models at the MSZ  using available seismic, geodetic, and historical catalogue data. Moreover, both aleatory and epistemic uncertainties were considered to obtain more accurate and reliable results.

The epistemic uncertainties were incorporated by combining event tree and ensemble modeling, including uncertainties of fault source and rupture complexity (dimensions, slip distribution, and possible locations of earthquakes).
The aleatory variability was identified from three main contributions (numerical model and bathymetry, tidal variation, and scaling relation), and incorporated directly into the probability equations, see~\eqref{Eq:aleatory}. 
Our results are demonstrated using hazard curves and probability maps. We also assessed the effect of the aleatory variability. To the best of our knowledge, this is the first PTHA sensitivity analysis concerning aleatory variability in the MSZ.
The findings are highlighted as follows.
\begin{enumerate}
	\item  The spread of hazard curves for different locations along the Makran coast is remarkably large. The probability that the tsunami height exceeds 3 m for return periods $\Delta T= \{50,$ $100, 250, 500, 1000\}$ ranges from $0$ to $\{13.5, 25, 52, 74, 91\}$ percent, respectively, for different PoIs.
	
	\item The probability of exceedance at PoIs near populated cities decreases and becomes insignificant for the exceedance threshold of 4 m (even for a long return period), except for Jask at the western coast of Iran. Our results provide evidence that if we consider the western part of the MSZ equally active and potential to the eastern part --similar to this paper, by weighting both parts the same in our event tree--the exceedance probability could be higher at the western part for a long return period. This can be clearly seen from the probability maps where the exceedance probability of 3 m fluctuates and becomes maximum at the western part of the MSZ.
	
	\item The inclusion of the aleatory variability has a significant effect on the probability of exceedance, and not including it mostly leads to a remarkable underestimation in the PTHA with a median of $10\%$ difference for all PoIs. This difference is underscored by increasing the return periods and reaches $40\%$ at somewhere for 1000-year return period in the presence and absence of the aleatory variability.
\end{enumerate}
Owing to Makran’s economical, geographical, and strategic importance, Iran approved a plan for developing the southern Makran Coast on December, 2016 titled ``Makran Sustainable Development.'' This plan, along with the drought occurring recently in the neighboring cities, has led to an inevitable migration toward the coastlines, with Chabahar exhibiting $10\%$ population rate growth last year and ranking among the highest population growth rates globally in 2019. Hence, our results are of vital for various stakeholders for developing and implementing tsunami risk mitigation activities and guiding risk-aware city planning.

This study is the first step toward comprehensive and reliable mitigation plans and activities; however, it is important to acknowledge its limitations. Tsunami sources beyond earthquakes were not considered in this study. Notably, the only tsunami induced with combination of earthquake and landslide in word history occurred in Makran in 1945. Moreover, in September 2013, a landslide was recorded immediately following an earthquake in the MSZ~\citep{september}. This highlights the need to consider landslides~\citep{rastgoftar2016study} and their combination with earthquakes in future PTHA studies. Moreover, we disregarded the dynamic interaction between tides and tsunami waves. This works for a tsunami wave with one isolated peak; however, it may lead to hazard underestimation when the tsunami has several peaks with significant heights.
Finally, the necessity and incorporation of high quality bathymetry and topography data is paramount, especially, for mapping tsunami inundation and vulnerable areas, which is the aim of our next study.
\section*{Acknowledgments }
	The computation was carried out using the computer resource offered under the category of General Projects by Research Institute for Information Technology, Kyushu University. We would like to thank Editage (www.editage.com) for English language editing and  Ports and Maritime Organization of Iran (PMO) for providing us with bathymetry data.
\section*{Conflicts of interests}
The authors have no conflicts of interest to declare that are relevant to the content of this article.
\bigskip

%%%%%%%%%%%%%%%%%%%%%%%%%%%%%%%%%%%%%%%%%%%%%%%
%%%%%%%%%%%%%%%%%%%%%%%%%%%%%%%%%%%%%%%%%%%%%%%
%%%%%%%%%%%%%%%%%%%%%%%%%%%%%%%%%%%%%%%%%%%%%%%

\bibliographystyle{spbasic}
\bibliography{Mybib}

\end{document}